\documentclass[
reprint, longbibliography,
superscriptaddress,
preprintnumbers,nofootinbib,
nobibnotes,
amsmath,amssymb,
aps, pra
floatfix,
]{revtex4-2}

\usepackage[normalem]{ulem}
\usepackage{gensymb}
\usepackage{upgreek}
\usepackage{mathrsfs}
\usepackage{comment}
\usepackage{graphicx}
\usepackage{dcolumn}
\usepackage{bm}
\usepackage{floatrow}
\usepackage{float}
\usepackage{url}
\usepackage[%
  colorlinks=true,
  urlcolor=blue,
  linkcolor=blue,
  citecolor=blue,
]{hyperref}
\usepackage{xcolor}
\usepackage{textcomp}
\usepackage{amsmath}
\usepackage{varwidth}
\usepackage{placeins}
\usepackage{gensymb}
\usepackage{siunitx}
\hypersetup{
    colorlinks,
    linkcolor={red!50!black},
    citecolor={blue!50!black},
    urlcolor={blue!80!black}
}
\usepackage[center]{titlesec}
  \titleformat{\section}
   {\centering\normalfont\fontsize{10}{10}\bfseries}{\thesection}{1em}{}
  \titlespacing{\section}{0pt}{12pt plus 4pt minus 2pt}{6pt plus 2pt minus 2pt}
    \titleformat{\subsection}
   {\centering\normalfont\fontsize{10}{10}\bfseries}{\thesubsection}{1em}{}
  \titlespacing{\subsection}{0pt}{12pt plus 4pt minus 2pt}{6pt plus 2pt minus 2pt}

\usepackage{tabularx}
\newcolumntype{C}{>{\centering\arraybackslash}X} 
\usepackage{multirow}

\newcommand{\unm}{Center for High Technology Materials and Department of Physics and Astronomy, University of New Mexico, Albuquerque, NM, USA}

\newcommand{\bra}[1]{\langle #1|}
\newcommand{\ket}[1]{|#1\rangle}

\usepackage{amsfonts, amsmath, amssymb, graphicx, xcolor, verbatim, lipsum, dsfont, bm, hyperref, makecell, comment, upgreek, graphicx, dcolumn, array}

\usepackage[export]{adjustbox}
\usepackage[mathlines]{lineno}
\usepackage[T1]{fontenc}

\begin{document}

\title{The impact of microwave phase noise on diamond quantum sensing}

\author{Andris Berzins$^{\mathsection}$}
\affiliation{\unm}

\author{Maziar Saleh Ziabari$^{\mathsection}$}
\affiliation{\unm}

\author{Yaser Silani}
\affiliation{\unm}

\author{Ilja Fescenko}
\affiliation{\unm}
\affiliation{University of Latvia, Riga, Latvia}

\author{Joshua T. Damron}
\affiliation{\unm}
\affiliation{Oak Ridge National Laboratory, Oak Ridge, TN, USA}

\author{John F. Barry}
\affiliation{Lincoln Laboratory, Massachusetts Institute of Technology, Lexington, MA, USA}

\author{Andrey Jarmola}
\affiliation{ODMR Technologies Inc., El Cerrito, CA, USA}
\affiliation{University of California-Berkeley, Berkeley, CA, USA}

\author{Pauli Kehayias}
\affiliation{Lincoln Laboratory, Massachusetts Institute of Technology, Lexington, MA, USA}
\affiliation{Sandia National Laboratories, Albuquerque, NM, USA}

\author{Bryan A. Richards}
\affiliation{\unm}

\author{Janis Smits}
\email{smitsjanis@gmail.com}
\affiliation{\unm}

\author{Victor M. Acosta}
\email{vmacosta@unm.edu}
\affiliation{\unm}

\renewcommand{\thefootnote}{}{\footnote{$\mathsection$ A.~Berzins and M.~Saleh~Ziabari contributed equally.\vspace{2mm}}}

\date{\today}

\begin{abstract}
Precision optical measurements of the electron-spin precession of nitrogen-vacancy (NV) centers in diamond form the basis of numerous applications. The most sensitivity-demanding applications, such as femtotesla magnetometry, require the ability to measure changes in GHz spin transition frequencies at the sub-millihertz level, corresponding to a fractional resolution of better than $10^{-12}$. Here we study the impact of microwave (MW) phase noise on the response of an NV sensor. Fluctuations of the phase of the MW waveform cause undesired rotations of the NV spin state. These fluctuations are imprinted in the optical readout signal and, left unmitigated, are indistinguishable from magnetic field noise. We show that the phase noise of several common commercial MW generators results in an effective $\rm pT\,s^{1/2}$-range noise floor that varies with the MW carrier frequency and the detection frequency of the pulse sequence. The data are described by a frequency-domain model incorporating the MW phase noise spectrum and the filter-function response of the sensing protocol. For controlled injection of white and random-walk phase noise, the observed NV magnetic noise floor is described by simple analytic expressions that accurately capture the scaling with pulse sequence length and the number of $\pi$ pulses. We outline several strategies to suppress the impact of MW phase noise and implement a version, based on gradiometry, that realizes a ${>}10$-fold suppression. Our study highlights an important challenge in the pursuit of sensitive diamond quantum sensors and is applicable to other qubit systems with a large transition frequency.
\end{abstract}

\maketitle

\section{Introduction}
Precision optical measurements of the electron-spin precession of nitrogen-vacancy (NV) centers in diamond form the basis of numerous applications, ranging from imaging biomagnetism~\cite{GLE2015,FES2019,WAN2019,MCC2020,WEB2021,KAZ2024} to nuclear magnetic resonance (NMR) spectroscopy~\cite{ASL2017,GLE2018,SMI2019,BRU2023}, gyroscopes~\cite{JAS2019,SOS2021,JAR2021}, femtotesla magnetometry~\cite{FES2020,SIL2023,GAO2023,BAR2023}, and searches for new spin physics~\cite{RON2018,CHU2022,WU2023}. As solid-state electron spin sensors, NV centers offer advantages over alkali-metal vapor and superconducting quantum interference device sensors in that a high density of immobile spins form a tunable sensing voxel that can be tailored for the application. Much attention has been devoted to the use of single or small ensembles of NV centers as nanoscale sensors, leading to remarkable advances in materials and biological microscopy~\cite{CAS2018,ZHA2021,MAR2022,ASL2023}. However, the development of ultra-sensitive bulk NV sensors, as needed for the most demanding metrology applications~\cite{TAY2008,ACO2009}, presents additional challenges. 

\begin{figure*}[bt]
      \includegraphics[width=0.99\textwidth]{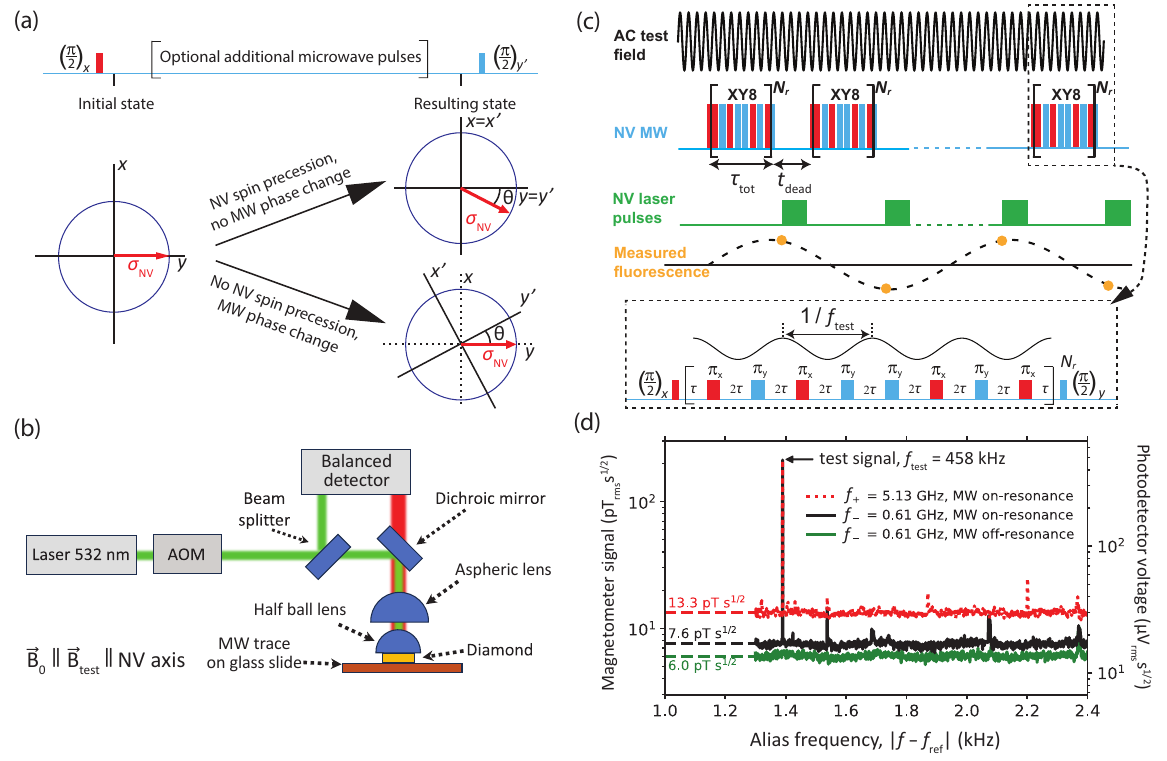}
    \caption{\textbf{MW phase noise detection: principle and experimental apparatus.} \textbf{(a)} Graphical representation of the impact of MW phase noise on a generic spin-$1/2$ sensor. MW phase errors lead to a rotation of the basis, with respect to the spin Bloch vector $\hat{\sigma}_{\rm nv}$, that is indistinguishable from spin precession due to a magnetic field. \textbf{(b)} Schematic of the NV magnetometer used for the measurements; detailed information can be found in \ref{app:specifics}. The bias field $\vec{B_0}$ and calibrated test fields $\vec{B}_{\rm test}$ are aligned to one of the NV axes (\ref{app:Bgeometry}).
    \textbf{(c)} Synchronized series of XY8-$N_r$ pulse sequences used for AC magnetic field measurements. \textbf{(d)} Typical magnetic spectra taken using MW generator G2. The right vertical axis is the processed photodetector voltage spectral density, and the left vertical axis is converted to magnetic sensitivity units.
    }
  \label{fig1}
\end{figure*}

One challenge is that NV electron spin transition frequencies are almost always large. Due to the zero-field splitting arising from embedding electron-spin $S=1$ defects in a solid, the magnetic-dipole-allowed NV spin transitions are ${\sim}3~{\rm GHz}$ at low magnetic field. A femtotesla-level NV sensor must measure changes in these spin transition frequencies at the sub-mHz level, corresponding to a fractional resolution substantially better than $10^{-12}$. In any spin-precession measurement, random phase fluctuations of the microwave (MW) control field lead to undesired rotations of the spin state that are often indistinguishable from magnetic field noise, see Fig.~\ref{fig1}(a). Phase noise is always present at some level, due to the limited clock stability and Johnson noise in MW signal generators~\cite{BAL2016,RUB2008}, but it has only recently been a limiting factor for NV precision measurements~\cite{SMI2019,SIL2023,BAR2023}. The magnitude of this effect should not be understated--phase noise from typical MW generators produces NV sensor noise at the level of $0.1{\mbox{-}}100~{\rm pT s^{1/2}}$, orders of magnitude above requirements for applications like magnetoencephalography~\cite{BAR2016,IBR2021,HAH2022,CHE2023,SEK2023}. Moreover, the impact of phase noise tends to grow with the MW frequency, which has implications for high-field applications like NMR chemical analysis.

Here, we experimentally characterize the impact of MW phase noise on the noise floor of an NV sensor. We show how to predict the impact numerically, using a frequency-domain model incorporating the MW phase noise spectrum and the filter-function response of the sensing protocol, and also with simple analytic expressions for the cases of white and random-walk phase noise. We discuss strategies to suppress the impact of MW phase noise and implement a version based on gradiometry that realizes a ${>}10$-fold suppression. Our results highlight a key factor in the design of sensitive diamond quantum sensors and are broadly applicable to other qubit systems.

\section{Experimental Setup}
\label{sec:exptsetup}
The principle behind the impact of MW phase noise is depicted in Fig.~\ref{fig1}(a). Consider a general Ramsey-type measurement on a spin $S=1/2$ qubit initially prepared in an $S_z$ eigenstate. A resonant MW $\pi/2$ pulse, with a well-defined phase, rotates the spin state to lie along the $y$-axis of the Bloch sphere in the rotating frame. In the first case, a small additional magnetic field is present, and the spin state precesses, accumulating a phase $\theta$ with respect to the $y$-axis. A final MW $\pi/2$ pulse is applied with the same frequency as the initial $\pi/2$ pulse but with a $90\degree$ phase shift. This pulse faithfully projects the phase accumulation onto the $S_z$ basis, and measurements of $\langle S_z\rangle$ can be used to determine $\theta$. In the second case, there is no additional magnetic field present, and the spin state does not precess. However, if the phase of the MW carrier of the final $\pi/2$ pulse has an error $\theta$ with respect to the desired phase, then the state projection and readout will produce the same result as in the first case. Thus, in this common scenario, MW phase noise is indistinguishable from magnetic field noise. This general concept holds even when there are additional MW pulses in the sequence, though the exact phase errors depend on the properties of the MW source and the pulse sequence.

Our experimental setup is depicted in Fig.~\ref{fig1}(b) and additional details are in~\ref{appendix:device}. A $120\mbox{-}{\rm \upmu m}$-thick, $(110)$-polished diamond membrane, with an NV density ${\sim}0.5~{\rm ppm}$, is adhered to a high-refractive-index ($n=2$) half-ball lens. Light from a $532\mbox{-}{\rm nm}$ laser passes through an acousto-optic modulator and is relayed by a $0.79{\mbox{-}}{\rm NA}$ aspheric lens onto the diamond membrane, resulting in an excitation beam waist of $50\mbox{-}100~{\rm \upmu m}$ and a power of ${\sim}0.4~{\rm W}$. NV center fluorescence is collected by the same aspheric lens, spectrally filtered (passing $650\mbox{-}800~{\rm nm}$ light), and focused onto one channel of a balanced photodetector. A small portion of the green excitation beam is directed to the second photodetector channel to suppress the impact of laser intensity fluctuations. The diamond together with the half-ball is adhered to a microscope slide that has a copper trace which delivers the MW field to the NV centers. A variable bias magnetic field $\vec{B_0}$, produced by an electromagnet, is aligned along one of the NV axes (\ref{app:Bgeometry}). The aligned NV centers have spin transition frequencies $f_{\pm}=D\pm\gamma_{\rm nv} B_0$, where $D=2.87~{\rm GHz}$ is the NV zero-field splitting and $\gamma_{\rm nv}=28.03~{\rm GHz/T}$ is the gyromagnetic ratio. A $55\mbox{-}{\rm mm}$-diameter wire loop is used to deliver uniform, calibrated (\ref{app:testcal}) oscillating (AC) test magnetic fields along the NV axis. Three commercial MW generators were studied, named G1, G2, and G3. Their model numbers and typical phase-noise performance are shown in Table~\ref{tbl:generator_table}.

\begin{table}[hbt]
\renewcommand{\arraystretch}{1.2}
\begin{tabular}{c | c | c} 
Name~ & ~MW generator model~ & Phase noise, $1~{\rm GHz}$\\ 
\hline
G1 & SRS SG386 & $-114~{\rm dBc/Hz}~(20~{\rm kHz})$ \\\hline
G2 & R\&S SMATE200A &  $-134~{\rm dBc/Hz}~(20~{\rm kHz})$  \\\hline
G3 & R\&S SMU200A-B22 & $-138~{\rm dBc/Hz}~(20~{\rm kHz})$ \\\hline
\end{tabular}
\caption{Microwave generators used in measurements. The final column is the manufacturer specifications of typical phase noise for a $1~{\rm GHz}$ carrier at $20~{\rm kHz}$ offset with I/Q modulation enabled~\cite{M_SRS386,M_SMATE200A}. Other information on phase-noise performance is in Fig.~\ref{fig2}(d)
and~\ref{appendix:Direct_PN}. 
}
\label{tbl:generator_table}
\end{table}

AC magnetometry is performed at room temperature using a synchronized series of XY8-$N_r$ pulse sequences~\cite{GLE2018,SMI2019,SIL2023}, see Fig.~\ref{fig1}(c). Each XY8-$N_r$ sequence (duration: $\tau_{\rm tot}$) begins and ends with a MW $\pi/2$ pulse that is resonant with one of the $f_{\pm}$ spin transitions. Between the $\pi/2$ pulses, a train of $8N_r$ resonant $\pi$ pulses (length: $t_{\pi}$), spaced by $2\tau = 1/(2f_{\rm xy8})-t_{\pi}$, are applied in a pattern of alternating phase. The XY8-$N_r$ pulse sequences are frequency selective, in that NV centers are primarily sensitive to AC magnetic field frequencies within a band, centered at $f_{\rm xy8}$, of width ${\sim}f_{\rm xy8}/(4N_r)$. Following each XY8-$N_r$ sequence, a $12\mbox{-}{\rm \upmu s}$ laser pulse is applied for NV optical readout and repolarization. The time between XY8-$N_r$ sequences, $t_{\rm dead}\approx15~{\rm \upmu s}$, accounts for this pulse as well as small additional delays. The sequence is repeated continuously, and the resulting time trace of NV fluorescence readouts is approximately proportional to an aliased version of an AC field that is sampled at the time of the first $\pi/2$ pulse of each XY8-$N_r$ sequence, with a sample rate $f_{\rm samp}=1/(\tau_{\rm tot}+t_{\rm dead})$. For an AC field of frequency $f$, that lies within the passband of the XY8-$N_r$ sequence, the NV fluorescence signal oscillates at an alias frequency $|f-f_{\rm ref}|$, where $f_{\rm ref}$ is the integer multiple of $f_{\rm samp}$ that is closest to $f$~\cite{SMI2019}.

Figure~\ref{fig1}(d) shows typical magnetic spectra taken using MW generator G2, with a $212~{\rm pT_{rms}}$ test field applied at $f_{\rm test}=f_{\rm xy8}=457.9~{\rm kHz}$ and $f_{\rm ref}=459.3~{\rm kHz}$. Here, $B_0=81~{\rm mT}$, so $f_+=5.13~{\rm GHz}$ and $f_-=0.61~{\rm GHz}$. To acquire a spectrum, a synchronized XY8-8 pulse series is applied continuously for ${\sim}150~{\rm s}$, the NV fluorescence time trace is split into $1\mbox{-}{\rm s}$ intervals, and the root-mean-squared average~\cite{GRA2024} of the absolute value of the Fourier transform of each interval is computed (\ref{appendix:processing}). Three spectra are shown: one with G2 tuned to the $f_+$ transition, one with G2 tuned to $f_-$, and one with G2 detuned by $+0.4~{\rm GHz}$ from $f_-$ for all MW pulses. The noise floor off resonance is $\eta_{\rm off}\approx6.0~{\rm pT_{rms}\,s^{1/2}}$, consistent with the expected noise floor in the photoelectron-shot-noise limit, $\eta_{\rm psn}\approx5.4~{\rm pT_{rms}\,s^{1/2}}$ (\ref{appendix:psn}). Measurements with G2 detuned by $+0.4~{\rm GHz}$ from $f_+$ have the same noise floor and are omitted for visual clarity. However, the spectrum for G2 on resonance with $f_{-}$ has a slightly higher noise floor, and the spectrum for G2 on resonance with $f_+$ has a more than two-fold higher noise floor. Assuming that the noise contributions are independent, we compute an excess noise as $\eta_{\rm ex,\pm}=(\eta_{\pm}^2-\eta_{\rm off}^2)^{1/2}$, where $\eta_{\pm}$ is the noise floor when MW are resonant with the $f_{\pm}$ transitions. The non-zero values, $\{\eta_{\rm ex,-}{\approx}4.7~{\rm pT_{rms}\,s^{1/2}},\eta_{\rm ex,+}{\approx}11.9~{\rm pT_{rms}\,s^{1/2}}\}$, indicate additional noise that is only present when G2 is tuned to resonance. This effect is reproducible regardless of the order of acquiring the spectra, the use of spectral filters in the MW chain, the type of MW amplifier used, or the method of phase alternation (\ref{appendix:earlytries}). However, the results change dramatically when using different MW generators, as will be described in Sec.~\ref{sec:freqGendependece}. That the noise floor would rise with increasing MW carrier frequency is an early hint that MW phase noise, which also tends to increase with carrier frequency~\cite{RUB2008}, is responsible for the excess noise.

\section{Phase noise spectroscopy}
\label{sec:freqGendependece}

\begin{figure*}[ht]
      \includegraphics[width=0.95\textwidth]{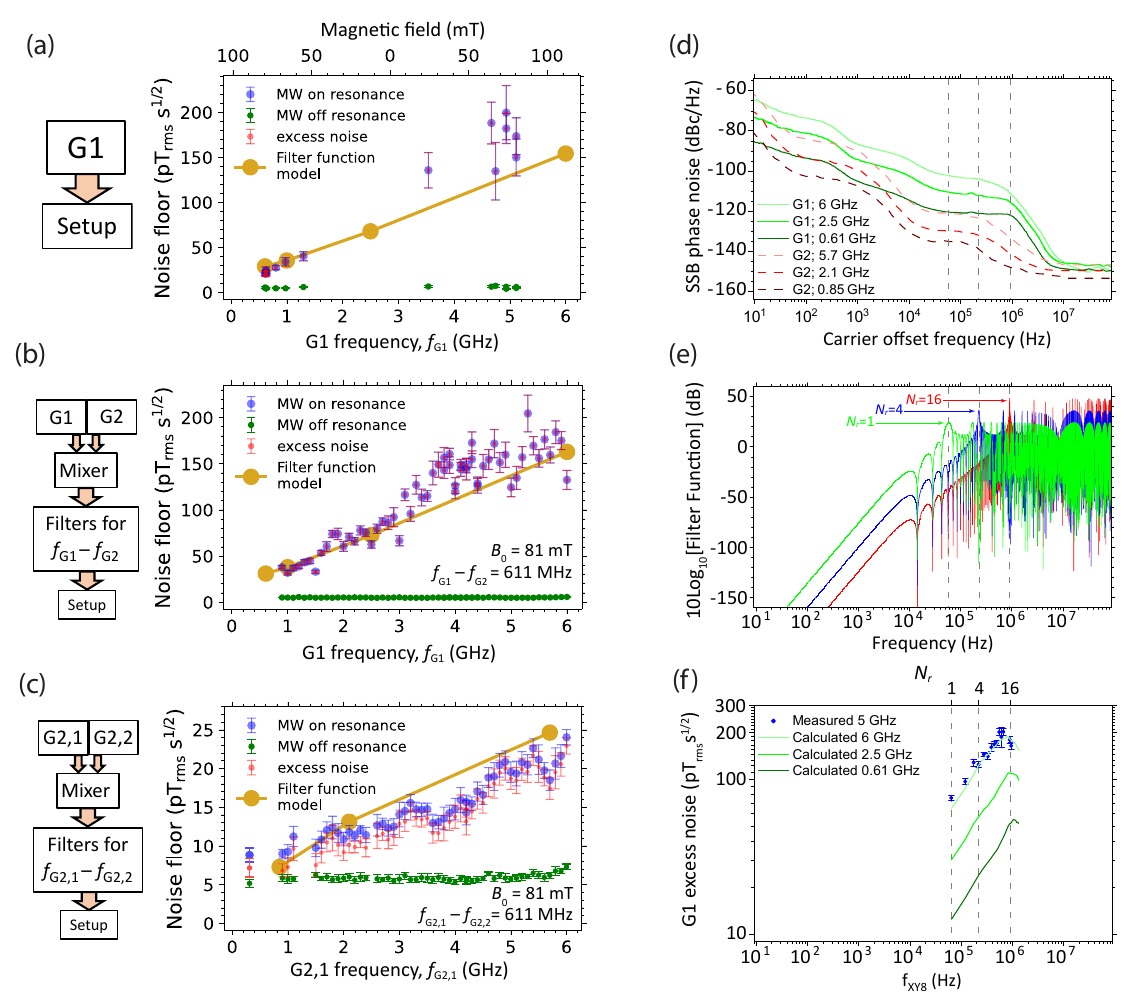}
    \caption{\textbf{Phase noise spectroscopy}
    \textbf{(a)} NV sensor noise floor for an XY8-6 pulse series with MW generator G1 tuned to a $f_{\pm}$ spin transition at different magnetic fields. For G1 carrier frequency $f_{\rm G1}<2.87~{\rm GHz}$, microwaves are tuned to the $f_-$ spin transition, and otherwise they are tuned to $f_+$. \textbf{(b)} NV sensor noise floor as a function of $f_{\rm G1}$. Here $B_0=81~{\rm mT}$ is constant, and the final MW tone probes the $f_-=611~{\rm MHz}$ transition. This tone is generated by jointly varying and mixing G1 and G2 carrier frequencies, while maintaining $f_{\rm G1}-f_{\rm G2}=611~{\rm MHz}$. \textbf{(c)} NV sensor noise floor as a function of $f_{\rm G2}$. Here, two independent channels of G2 are varied and mixed, while maintaining $f_{\rm G2,1}-f_{\rm G2,1}=611~{\rm MHz}$. The lowest frequency represents $f_{\rm G2,1}+f_{\rm G2,2}= 611~{\rm MHz}$ (\ref{app:specifics}). \textbf{(d)} G1 and G2 single-sideband (SSB) phase noise power spectra, $\mathcal{L}(f)$, for different MW carrier frequencies. For G1, the spectra are measured using a phase-noise analyzer,~\ref{appendix:Direct_PN}. For G2, they are extracted from the manufacturer specification sheet \cite{M_SMATE200A}. \textbf{(e)} XY8-$N_r$ filter functions, $\mathcal{F}(f)$, for $N_r=\{1,4,16\}$. Dashed gray lines are aligned with the first harmonic, $f_{\rm XY8}$, of each function and are extended to (d) and (f) for comparison. \textbf{(f)} Excess noise, $\eta_{\rm ex,+}$, for G1 obtained for different values of $N_r$. Here $B_0=76~{\rm mT}$, $f_{\rm G1}=f_+=5.00~{\rm GHz}$, and $\tau_{\rm tot}\approx70~{\rm \upmu s}$ (\ref{appendix:processing}, \ref{appendix:filterfunc}). Varying $N_r$ also varies the filter function's first harmonic, $f_{\rm XY8}$. Green curves in (f), and gold points in (a-c), are calculated from the filter-function model, using Eqs.\eqref{eq:FF},\eqref{eq:etaphi} and curves in (d,e). For (a-c,f), error bars are the standard deviation of ${\geq}10$ identical measurements (\ref{appendix:processing}). 
    }
  \label{fig2}
\end{figure*}

To study the excess noise in more detail and verify its origin, we measured the NV sensor noise floor as a function of MW carrier frequency and detection frequency, $f_{\rm xy8}$, for two different MW generators--G1 and G2. Figure~\ref{fig2}(a) is a plot of the NV sensor noise floor as a function of G1 carrier frequency, $f_{\rm G1}$, under an XY8-6 pulse series. To generate this plot, $B_0$ is varied in the range $20\mbox{-}80~{\rm mT}$. For a given value of $B_0$, the noise floor is measured for G1 tuned to a $f_{\pm}$ resonance, as well as the off-resonance case (detuning: $0.2~{\rm GHz}$), and the excess noise $\eta_{{\rm ex},\pm}$ is inferred. Throughout, we observe markedly larger values of $\eta_{{\rm ex},\pm}$ for G1 than those observed with G2 [see Fig.~\ref{fig1}(d)] at the same carrier frequency. This is consistent with the higher phase noise of G1, as specified by the manufacturer (Table~\ref{tbl:generator_table}). Furthermore, we observe a roughly monotonic increase in excess noise with increasing MW generator frequency. A similar trend is observed for generator G2 (\ref{app:specifics}) when sweeping $B_0$ and matching $f_{\rm G2}$ to the $f_{\pm}$ resonances.

In order to isolate the MW carrier-frequency dependence, and eliminate any effects due to magnetic field, we next fixed the magnetic field at $B_0=81~{\rm mT}$. We mixed the output of G1 with one of the output channels of (low-noise) G2 and filtered for the difference frequency, $f_{\rm G1}-f_{\rm G2}$. Since G2 has a much lower phase noise (Table~\ref{tbl:generator_table}), we assume the phase noise of the difference frequency is dominated by the G1 phase noise (\ref{app:mixer}). For MW-on-resonance measurements, we set $f_{\rm G1}-f_{\rm G2}=f_-=611~{\rm MHz}$, and for the off-resonance case we set $f_{\rm G1}-f_{\rm G2}=811~{\rm MHz}$. This allowed us to vary $f_{\rm G1}$ without altering the NV properties. Figure~\ref{fig2}(b) shows the measured NV sensor noise floors, under an XY8-8 pulse series, as a function of $f_{\rm G1}$ in the $0.9{\mbox{-}}6~{\rm GHz}$ range. The behavior is similar to that of Fig.~\ref{fig2}(a), reinforcing the monotonic dependence of $\eta_{\rm ex,\pm}$ on MW carrier frequency.

We used a similar method to probe the NV sensor noise dependence on $f_{\rm G2}$, Fig.~\ref{fig2}(c). In this case, two output channels of G2 are mixed. We expect that the phase noise of each channel is uncorrelated, and thus the phase noise of the difference frequency is ${\sim}\sqrt{2}$ times larger than that of a single channel (\ref{app:mixer}). The measured excess noise increases roughly monotonically with $f_{\rm G2,1}$, but it is ${\sim}7$ times lower than that observed in Fig.~\ref{fig2}(b). This is qualitatively consistent with the lower phase noise for G2 compared to G1. 

To predict the impact of MW generator phase noise on the NV sensor noise floor, we apply a frequency-domain, filter-function method~\cite{BAL2016}. The method incorporates the single-sideband power spectral density, $\mathcal{L}(f)$, for G1 and G2 at different carrier frequencies, as shown in Fig.~\ref{fig2}(d) (see also \ref{appendix:Direct_PN}). The standard deviation of the NV spin state's phase displacement, $\sigma_{\phi}$, due to MW phase errors in a single pulse sequence is given by (\ref{appendix:filterfunc}):
\begin{equation}
\label{eq:FF}
\sigma_{\phi}^{2}=\int_{0}^{\infty}S_{\phi}(f)\,\mathcal{F}(f)\,df\approx\int_{0}^{f_c}S_{\phi}(f)\,\mathcal{F}(f)\,df,
\end{equation}
where $S_{\phi}(f)=2\times10^{\frac{\mathcal{L}(f)}{10}}$, $\mathcal{F}(f)$ is the filter function of the multipulse sequence, and $f_c\approx0.1~{\rm GHz}$ is a cutoff frequency that depends on the frequency response of the MW delivery (\ref{app:cutoff}). Expressions for $\mathcal{F}(f)$ of an XY8-$N_r$ sequence~\cite{Bie2011} are given in \ref{appendix:filterfunc}, and the cases of $N_r=1$, 4, and 16 are shown in Fig.~\ref{fig2}(e). 

The impact of MW phase noise results in an equivalent magnetic sensitivity given by: 
\begin{equation}
\label{eq:etaphi} \eta_{\phi}\approx\frac{\sigma_{\rm \phi}}{4\,\gamma_{\rm nv}\,\sqrt{\tau_{\rm tot}}}\,\sqrt{1+t_{\rm dead}/\tau_{\rm tot}}.
\end{equation}
This noise contribution is uncorrelated with that due to photoelectron shot noise and can thus be directly compared to $\eta_{\rm ex,\pm}$. In Figs.~\ref{fig2}(a-c), the noise floors calculated from $\eta_\phi$ [Eqs.~\eqref{eq:FF} and \eqref{eq:etaphi}] are shown, using the same pulse sequence parameters as in the experiment (see \ref{appendix:processing} and \ref{appendix:filterfunc}). For higher MW carrier frequencies, the calculated and experimental values differ slightly, perhaps due to $1\mbox{-}2~{\rm dB}$ disparities in $\mathcal{L}(f)$. However, overall, the calculated values are largely in agreement with the measurements, indicating that the source of $\eta_{\rm ex}$ is indeed from MW phase noise.

The filter-function model of Eqs.~\eqref{eq:FF} and \eqref{eq:etaphi} implies that, for a given MW signal generator, the NV sensor noise floor depends on the choice of pulse sequence. To probe this effect, we perform experiments where we fix $\tau_{\rm tot}\approx70~{\rm \upmu s}$ (the experimental $\tau_{\rm tot}$ values vary slightly, but for calculations we use the median value $\tau_{\rm tot}=70~{\rm \upmu s}$) and vary $N_r$ (\ref{appendix:processing}, \ref{appendix:filterfunc}). Here $B_0{=}76~{\rm mT}$, and G1 probes the $f_+{=}5~{\rm GHz}$ transition. 

Figure~\ref{fig2}(f) shows the resulting measured values of $\eta_{\rm ex,+}$, along with the noise floors calculated from $\eta_{\phi}$ [Eqs.~\eqref{eq:FF} and \eqref{eq:etaphi}] for different G1 carrier frequencies. While a comparison under exactly the same carrier frequency was not possible, the experimental and simulated curves have the same shape. We can qualitatively understand the shape as follows. The G1 phase noise spectrum, $S_{\phi}(f)$, is approximately flat for carrier offset frequencies between $50\mbox{-}500~{\rm kHz}$ (white noise band), and then it falls sharply as ${\sim}1/f^2$ for offset frequencies between $0.5\mbox{-}5~{\rm MHz}$ (random-walk noise band). When the NV sensor detection frequency falls within the white noise band ($f_{\rm xy8}\lesssim500~{\rm kHz}$, $N_r\lesssim9$), the scaling of the NV sensor excess noise with $N_r$ is dominated by the scaling of the integral of $\mathcal{F}(f)$ within this band, see Eq.~\eqref{eq:FF}. For a sequence of periodic $\pi$ pulses, including XY8-$N_r$, the peak heights grow as $N_r^2$, while the number of harmonics that fall within the band shrinks as $1/N_r$. Thus, $\sigma_{\phi}^2$ grows approximately as $N_r$ and $\eta_{\phi}$ grows approximately as $\sqrt{N_r}$. However, when the NV sensor detection frequency falls within the random-walk noise band ($f_{\rm xy8}\lesssim500~{\rm kHz}$, $N_r\lesssim9$), the $N_r^2$ increase in $\mathcal{F}(f)$ peak heights is compensated for by the ${\sim}1/f^2$ decrease in $S_{\phi}(f)$, leading to a plateau in excess noise and even a slight decrease when $N_r\gtrsim14$.

Based on these results, we conclude: i) noise spectroscopy using a multipulse NV sensor carries information on the MW generator phase noise spectrum that can be inferred through spectral decomposition methods~\cite{ALV2011,BYL2011,BAR2012}, and ii) the choice of pulse sequence has a large impact on the phase-noise-limited NV sensor noise floor and is thus an important consideration in the design of NV precision experiments. 

In \ref{app:cwodmr}, we also analyze the phase-noise response of a continuous-wave optically-detected magnetic resonance measurement. In this case, the phase-noise-limited equivalent magnetic sensitivity tends to be better than that for multipulse sequences, and it depends on the magnetometer bandwidth. However, the impact of phase noise is still important. For example, for a $1{\mbox{-}}{\rm kHz}$ bandwidth magnetometer, the phase-noise-limited equivalent magnetic sensitivity is ${\sim}1.4~{\rm pT_{rms}\,s^{1/2}}$ for G1 operated at $f_{\rm G1}=2.5~{\rm GHz}$ and ${\sim}0.3~{\rm pT_{rms}\,s^{1/2}}$ for G2 operated at $f_{\rm G2}=2.1~{\rm GHz}$.

\section{White and random-walk phase noise}
\label{Noise_injection}

\begin{figure*}[hbt]
\includegraphics[width=0.99\textwidth]{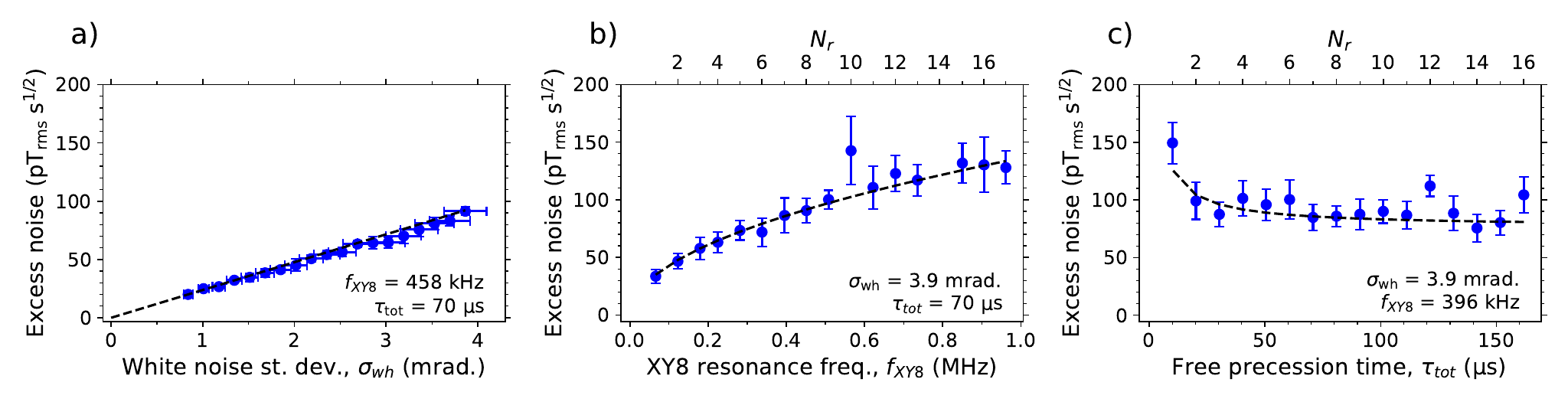}
    \caption{\textbf{White phase-noise injection.} Blue dots are measured values. Black curves are calculated from Eq.~\eqref{eq:etaW} using independently-measured parameters (\ref{appendix:processing}). 
    \textbf{(a)} NV sensor excess noise as a function of the standard deviation of MW phase errors between $\pi$ pulses, $\sigma_{\rm {wh}}$. \textbf{(b)} Excess noise versus $N_r$ (and thus $f_{\rm xy8}$), holding $\tau_{\rm tot}$ and $\sigma_{\rm wh}$ constant. \textbf{(c)} Excess noise versus $N_r$ (and thus $\tau_{\rm tot}$), holding $f_{\rm xy8}$ and $\sigma_{\rm wh}$ constant. In equivalent magnetic noise calculations, we used $t_{\rm dead} = 17~{\rm \upmu s}$. In (a-c), error bars are the standard deviation of ${\geq}10$ identical measurements. We used $t_{\rm dead}{=}17~{\rm \upmu s}$ for magnetic noise calculations.
    }
  \label{fig4}
\end{figure*}

\begin{figure*}[htb]
      \includegraphics[width=0.99\textwidth]{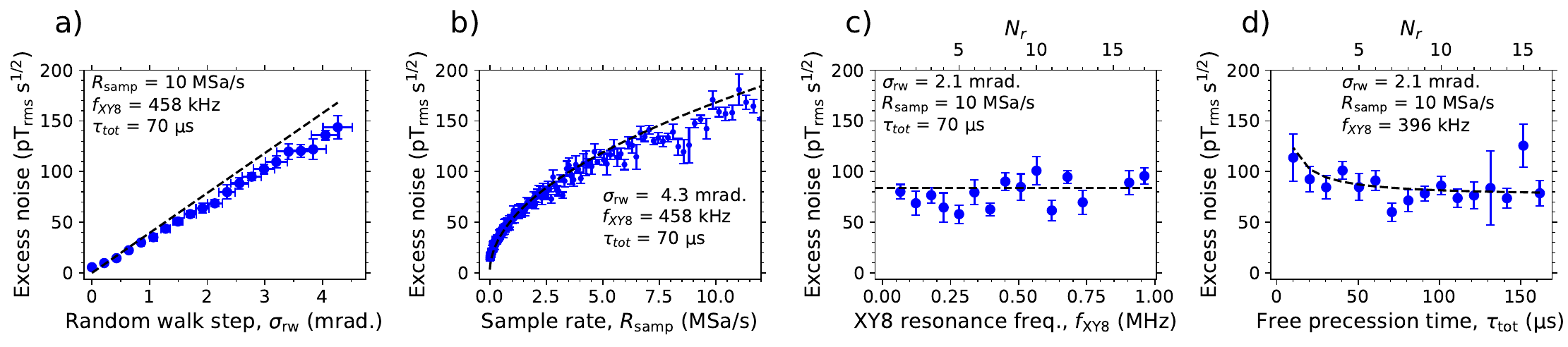}
    \caption{\textbf{Random-walk phase-noise injection.} Blue dots are measured values. Black curves are calculated from Eq.~\eqref{eq:M_SenNoisRW} using independently-measured parameters (\ref{appendix:processing}). \textbf{(a)} NV sensor excess noise versus the standard deviation of random-walk phase jumps $\sigma_{\rm rw}$. \textbf{(b)} Excess noise versus jump sample rate, $R_{\rm samp}$. \textbf{(c)} Excess noise versus $N_r$ (and thus $f_{\rm xy8}$), holding $\tau_{\rm tot}$, $R_{\rm samp}$, and $\sigma_{\rm rw}$ constant. \textbf{(d)} Excess noise versus $N_r$ (and thus $\tau_{\rm tot}$), holding $f_{\rm xy8}$, $R_{\rm samp}$, and $\sigma_{\rm rw}$ constant. In (a-c), vertical error bars are the standard deviation of ${\geq}5$ identical measurements, and we used $t_{\rm dead} = 17~{\rm \upmu s}$ for magnetic noise calculations.
    } 
  \label{fig3}
\end{figure*}

We next detail the response of an NV sensor under the controlled injection of MW phase noise. We focus on the common cases of white and random-walk phase noise, for which we derive simple analytic formulas of their impact on an NV pulsed sensor (\ref{app:tdomainmodel}). Here, it is assumed that the MW carrier phase of a given pulse is well-defined, but the relative phase changes from pulse to pulse due to phase noise. 

For white phase noise, the standard deviation of the NV spin state's phase displacement due to MW phase errors in a single XY8-$N_r$ pulse sequence is $\sigma_{\phi}=2\sigma_{\rm wh}\sqrt{N+1/4}\approx2\sigma_{\rm wh}\sqrt{N}$, where $\sigma_{\rm wh}$ is the standard deviation of the MW phase error of each pulse, and $N=8N_r$ is the total number of $\pi$ pulses. Using Eq.~\eqref{eq:etaphi}, the equivalent NV sensor magnetic sensitivity is then:
\begin{equation}
\label{eq:etaW}
\eta_{\rm wh}\approx\frac{\sigma_{\rm wh}}{\gamma_{\rm nv}}\sqrt{\frac{f_{\rm xy8}}{2}}\sqrt{1+t_{\rm dead}/\tau_{\rm tot}}.
\end{equation}

We experimentally validated the behavior of the NV sensor noise floor under white MW phase noise, by injecting pseudo-white noise into the phase modulation port of G3 (\ref{app:noise_prep}). Here, $B_0{=}81~{\rm mT}$ and $f_{\rm G3}{=}f_{-}{=}0.61~{\rm GHz}.$ Figure~\ref{fig4} shows the NV sensor excess noise as a function of (a) $\sigma_{\rm wh}$, (b) $f_{\rm xy8}$, and (c) $\tau_{\rm tot}$. In each case the experimental values of $\eta_{\rm ex}$ agree well with those calculated from Eq.~\eqref{eq:etaW}. Interestingly, in the MW phase noise limit, the noise floor grows as a function of $f_{\rm xy8}$ (for fixed $\tau_{\rm tot}$). This scaling is quite different from the scaling in the photoelectron-shot-noise limit (\ref{appendix:psn}), where the NV noise floor decreases due to improved contrast when the ``dynamical decoupling'' property of the multipulse sequence improves the NV coherence time, $T_2$. Furthermore, the MW-phase-noise-limited noise floor hardly changes as $\tau_{\rm tot}$ increases (for fixed $f_{\rm xy8}$). This is also a departure from the photoelectron-shot-noise limit, where an increase in $\tau_{\rm tot}$ should decrease the noise floor for $\tau_{\rm tot}\lesssim T_2$. 

The behavior of the NV sensor noise floor with white phase noise is particularly important, since the white Johnson noise of a MW oscillator sets a fundamental limit on a MW generator's phase-noise performance~\cite{HAT2003,RUB2008}. As discussed in~\ref{appendix:filterfunc}, for a $0~{\rm dBm}$ MW oscillator with a noise temperature of $300~{\rm K}$, $\mathcal{L}\approx-177~{\rm dBc/Hz}$, and the phase-noise-limited equivalent magnetic sensitivity would be at the ${\sim}100~{\rm fT\,s^{1/2}}$ level. This is far above fundamental limits set by photoelectron shot noise or spin projection noise~\cite{TAY2008}. In \ref{app:cwodmr}, we also analyze the phase-noise response of a continuous-wave optically-detected magnetic resonance measurement. We find that the NV sensor equivalent magnetic sensitivity in the oscillator Johnson noise limit is ${\gtrsim}10~{\rm fT\,s^{1/2}}$ for a magnetometer bandwidth $\gtrsim100~{\rm kHz}$.

For random-walk phase noise, the standard deviation of the NV spin state's phase displacement due to MW phase errors in a single XY8-$N_r$ pulse sequence is $\sigma_{\phi}=\sigma_{\rm rw}\,\sqrt{\tau_{\rm tot} R_{\rm samp}}$, where $\sigma_{\rm rw}$ is the standard deviation of MW phase jumps and $R_{\rm samp}$ is the jump rate. In deriving this expression (\ref{app:tdomainmodel}), we assumed that $R_{\rm samp}\gtrsim1/(2\tau)$ so the MW phase changes for each $\pi$ pulse. The equivalent NV sensor magnetic sensitivity is:
\begin{equation}
\label{eq:M_SenNoisRW} \eta_{\rm rw}\approx\frac{\sigma_{\rm rw}\sqrt{R_{\rm samp}}}{4\gamma_{\rm nv}}\sqrt{(1+t_{\rm dead}/\tau_{\rm tot})}.
\end{equation}

We injected pseudo-random-walk phase noise using G3 (\ref{app:noise_prep}) and observed the NV noise floor behavior experimentally ($B_0{=}81~{\rm mT}$, $f_{\rm G3}{=}f_{-}{=}0.61~{\rm GHz}$). Figure~\ref{fig3} shows the NV sensor excess noise as a function of (a) $\sigma_{\rm rw}$, (b) $R_{\rm samp}$, (c) $f_{\rm xy8}$, and (d) $\tau_{\rm tot}$. In each case, the experimental values of $\eta_{\rm ex}$ match those calculated from Eq.~\eqref{eq:M_SenNoisRW}, aside from minor deviations due to imperfect delivery of random-walk noise in the experiment (\ref{app:noise_prep}). Unlike the case of white phase noise, with random-walk noise, $\eta_{\rm ex}$ is independent of the number of $\pi$ pulses or $f_{\rm xy8}$. However, as with white noise (but unlike in the photoelectron-shot-noise limit) $\eta_{\rm ex}$ hardly changes with $\tau_{\rm tot}$. The case of random-walk phase noise is especially relevant for qubit sensors with a long coherence time, as MW generators are often limited by random-walk phase noise at offset frequencies $\lesssim10~{\rm kHz}$, see, for example, the behavior of G2 in Fig.~\ref{fig2}(d).

\section{Phase noise cancellation}
While MW phase noise represents an important technical challenge for NV precision measurements, various common-mode noise rejection methods can suppress its impact. Here, we implement a simple way to minimize the impact of MW phase noise using two-point gradiometry. The excitation laser beam is split into two beams that are focused to separate spots in the same diamond, located on opposite sides of the MW trace (\ref{app:gradiometry} and \ref{app:Bgeometry}). The emission from each spot is directed to separate channels of the balanced photodetector. The detection spots are close enough apart (${\sim}0.2~{\rm mm}$) that NV centers in each spot are subject to approximately the same MW field (aside from an overall ${\sim}180\degree$ phase difference), and thus similar phase noise. In these experiments, the diamond is rotated within the electromagnet to interrogate one of the out-of-plane NV axes. 

\begin{figure}[hbt]
    \includegraphics[width=0.91\columnwidth]{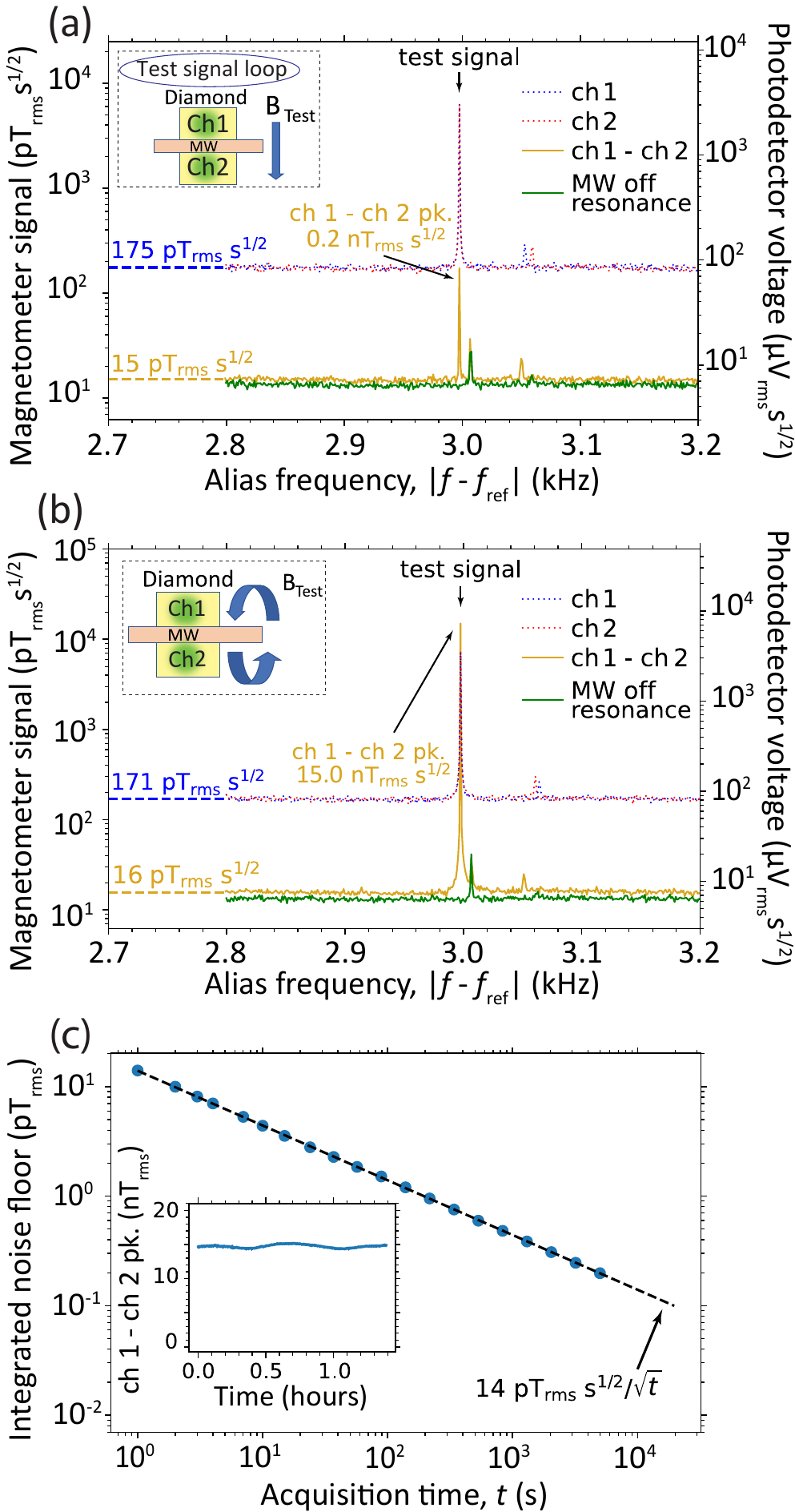}
    \caption{\textbf{Phase noise cancellation by gradiometry.} G1 is used with a XY8-7 series at $B_0{=}76~{\rm mT}, f_+{=}5~{\rm GHz}, f_{\rm test}=394~{\rm kHz}$~(\ref{appendix:processing}). \textbf{(a)} Magnetometer spectra obtained with a uniform test signal. Blue and red curves are spectra for individual channels (ch 1 and ch 2), corresponding to emission from each spot separately. The gradiometer difference signal is in yellow. The MW-off-resonance noise for the gradiometer is in green. \textbf{(b)} Magnetometer spectra obtained for a gradient test signal. The gradiometer signal shows a peak that is nearly twice that of individual channels. The noise floor is suppressed by a factor of ${\sim}11$ and is near the MW-off-resonance limit. \textbf{(c)} Gradiometer integrated noise floor as a function of acquisition time, $t$. The data are well fit by a $14~{\rm pT_{rms}\,s^{1/2}}/\sqrt{t}$ curve out to $t=5000~{\rm s}$. (inset) Stability of the gradiometer test signal amplitude over $5000~{\rm s}$.
    } 
  \label{fig5}
\end{figure}

Magnetometer spectra were obtained using G1 with an XY8-7 pulse series under two applied test field configurations. In the first configuration, Fig.~\ref{fig5}(a), a $394~{\rm kHz}$ test field is generated from a large coil (\ref{appendix:device}), resulting in a uniform test magnetic field over the two spots. The magnetic spectra of each spot separately (emission from the other channel was blocked) feature a $6.3~{\rm nT_{rms}}$ test-signal peak, and the MW-phase-noise-limited noise floor is ${\sim}175~{\rm pT_{rms}\,s^{1/2}}$. However when operating as a gradiometer (neither channel is blocked), the test signal is highly suppressed to $\lesssim0.2~{\rm nT_{rms}}$. Moreover, the noise floor drops to $15~{\rm pT_{rms}\,s^{1/2}}$, a factor of ${\sim}12$ suppression compared to the single-spot case. 

In the second configuration, a $394~{\rm kHz}$ gradient test field is applied by sending the test signal through the same trace on the chip as the MW radiation, Fig.~\ref{fig5}(b). Here, magnetic spectra of each spot separately exhibit a ${\sim}7.5~{\rm nT_{rms}}$ test-signal peak, with a MW-phase-noise-limited noise floor of ${\sim}170~{\rm pT_{rms}\,s^{1/2}}$. However, when operating as a gradiometer (neither channel is blocked), the test signal peak is approximately doubled to $15~{\rm nT_{rms}}$, which is expected since the sign of the test field is opposite between the two spots. Nevertheless, the gradiometer noise floor drops to $16~{\rm pT_{rms}\,s^{1/2}}$, an ${\sim}11$-fold suppression compared to the single-spot case.

We tested the stability of the gradiometer as a function of acquisition time, $t$, over the course of $5000~{\rm s}$. The noise floor exhibits a $14~{\rm pT_{rms}\,s^{1/2}/\sqrt{t}}$ scaling for the duration of the measurement, dropping below $200~{\rm fT_{rms}}$ for $t=5000~{\rm s}$. Moreover, the gradient signal peak remains at $14.7\pm0.5~{\rm nT_{rms}}$ over the course of the measurement, indicating a stable scale factor.

\section{Discussion}
\label{sec:discuss}
Two-point gradiometry is a powerful method for phase-noise suppression when detecting localized fields that vary substantially over millimeter length scales. In cases where fields vary less sharply, or when only a single detection spot is desired, the use of multiple NV transitions can be exploited. 

For example, for a small applied field along the NV axis ($B_0\ll D/\gamma_{\rm nv}=0.1~{\rm T}$), a double-quantum pulse sequence can be applied to suppress the impact of MW phase noise (\ref{app:dq}). Two MW tones at $f_{\pm}$ can be generated by mixing a large carrier frequency ($f_{\rm car}=D$) with a much lower-frequency local oscillator ($f_{\rm LO}=\gamma_{\rm nv} B_0$)~\cite{MAM2014}. As derived in \ref{app:dq}, the resulting NV sensor signal then depends only on the phase noise of the local oscillator, and it is independent of the phase noise of the $f_{\rm car}$ source. If the phase noise varies linearly with carrier frequency, as is approximately the case for the MW generators studied here, this scheme allows for a suppression of the MW phase noise impact by a factor of $f_{\rm car}/f_{\rm LO}$. 

For a continuous-wave NV measurement, at low magnetic field, the same mixer scheme can be used with a dual-resonance technique~\cite{WOJ2018,FES2020} to realize comparable levels of suppression of the MW phase noise impact (\ref{app:cwdual}). 

At very low applied field $B_0\lesssim\Omega_R/\gamma_{\rm nv}$, where $\Omega_R$ is the MW Rabi frequency, a single MW tone can be used with a double-quantum pulse sequence~\cite{FAN2013}, potentially eliminating the first-order impact of MW phase noise entirely.

In some applications, interrogating both NV resonances may not be feasible or effective. In either case, pulse sequences may be improved by incorporating MW phase noise into the optimization of pulse-sequence filter functions. The dependence of the NV phase-noise-limited sensitivity on the filter-function detection frequency, Fig.~\ref{fig2}(f), points to the possible efficacy of this strategy. Pulse sequences that take advantage of low-phase-noise regions of a generator’s spectrum (for example, due to the behavior of the phase-locked loop), or the use of composite, chirped, or aperiodic pulses, may offer superior performance. When practical, ultra-low phase noise MW generators based on superconducting resonators~\cite{CHA2020}, dielectric cavities~\cite{IVA2009,HAT2014}, ferrimagnetic oscillators~\cite{GOR2014,BAR2023YIG}, and photonic microwave generators~\cite{YAO1996,LI2023,KUD2024} may be helpful. Finally, the use of magnetic flux concentrators~\cite{FES2020,SHA2023,GAO2023,SIL2023} provides some relief from the impact of phase noise, as magnetic signals are amplified while the impact of phase noise is unchanged. 

In summary, we studied the impact of MW phase noise on NV sensors and verified that it is an important limiting factor for high-sensitivity experiments. We showed how to quantitatively predict the impact of MW phase noise, with knowledge of a generator's phase noise spectrum and the pulse sequence filter function. We provided simple analytic expressions for the case of white and random-walk phase noise. Finally, we identified a number of methods to suppress the impact of phase noise and implemented one, based on two-point gradiometry, that provides a >10-fold suppression. Our results inform on the design of precision measurements and spectroscopy experiments using NV centers and other qubit systems with a large transition frequency.

\begin{acknowledgments}
We gratefully acknowledge advice and support from P.~Schwindt, D.~Thrasher, T.~Drake, C.~Ramanathan, I.~Savukov, D.~Budker, P.~R.~Hemmer, A.~Jeronimo~Perez, and anonymous reviewers. We especially thank A.~Orozco for loaning and assisting with measurements using a MW phase-noise analyzer.\\
\textbf{Competing interests.} The authors declare no competing financial interests.\\
\textbf{Author contributions.} J.~S., A.~B., M.~S.~Z., and V.~M.~A. conceived the idea and designed the experiments. A.~B. and M.~S.~Z. built the main experimental apparatus, acquired and analyzed data, and wrote the initial manuscript draft. J.~T.~D., I.~F., Y.~S., and J.~S. made early measurements using previous setups. J.~F.~B., P.~K., A.~J., Y.~S., B.~A.~R., and J.~S. assisted in theoretical modeling, experimental execution, and data interpretation. M.~S.~Z. and J.~S., wrote the control software. V.~M.~A. supervised the project. All authors helped edit the manuscript. \\
\textbf{Funding.} This work was supported by the National Science Foundation (CHE-1945148, OIA-1921199) and National Institutes of Health (DP2GM140921, R41GM145129).

\end{acknowledgments}

\clearpage
\appendix
\setcounter{equation}{0}
\setcounter{section}{0}
\makeatletter
\renewcommand{\thetable}{A\arabic{table}}
\renewcommand{\theequation}{A\Roman{section}-\arabic{equation}}
\renewcommand{\thefigure}{A\arabic{figure}}
\renewcommand{\thesection}{Appendix~\Roman{section}}
\makeatother

\section{Experimental setup details}
\label{appendix:device}
 Our diamond sensor is similar to those used in Refs.~\cite{SMI2019,SIL2023}. The diamond was grown by chemical vapor deposition, yielding an initial nitrogen density of ${\sim}5~{\rm ppm}$. The diamond was irradiated with 2-MeV electrons at a dose of $3\times10^{17}~{\rm cm^{-2}}$, and it was subsequently annealed at $800^{\circ}$ and $1100^{\circ}~{\rm C}$ using the recipe described in Ref.~\cite{KEH2017}. Based on fluorescence brightness, we estimate the resulting NV$^-$ concentration is ${\sim}0.5~{\rm ppm}$. Finally, the diamond was cut into membranes of dimensions ${\sim}800\times600\times120~{\rm \upmu m^3}$ with a (110) surface polish. 

The diamond sensor is sandwiched between a glass slide with a copper trace and a half-ball lens ($8~{\rm mm}$ diameter, refractive index $n=2$). A Norland Optical Adhesive 88 is used at the interfaces to mechanically secure the diamond in place and improve optical contact. Microwave fields are delivered to the NV centers through the ${\sim}2\mbox{-}{\rm \upmu m}$-thick, lithographically-defined copper trace, which has a trace width of ${\sim}200~{\rm \upmu m}$ in the region in contact with the diamond. The laser light ($532~{\rm nm}$, Lighthouse Photonics Sprout-G-10W) passes through an acousto-optic modulator (Brimrose TEM-110-25-532) to form $12~{\rm \upmu s}$ pulses with a peak power ${\sim}0.4~{\rm W}$. The light subsequently passes through an additional lens [not shown in Fig.~\ref{fig1}(a) of the main text] and is relayed onto the diamond using a ${\rm NA}=0.79$ aspheric lens. The arrangement of lenses (including the $n=2$ half-ball lens) is selected to provide wide-field illumination with an excitation diameter in the range $50{\mbox{-}}100~{\rm \upmu m}$. NV fluorescence is collected by the same lens, is spectrally filtered by a dichroic mirror (Thorlabs DMLP550L) and longpass filter (Thorlabs FELH0650), and is directed to one channel of a balanced photodetector (Thorlabs PDB210A) with a transimpedance gain of $G{=}175~{\rm kV/A}{=}2.8{\times}10^{-14}~{\rm V/(photoelectron/s)}$ at $50~\ohm$ impedance. 

Considering the ${\sim}50~{\rm \upmu m}$ excitation beam diameter, the ${\sim}120~{\rm \upmu m}$ diamond membrane thickness, and the ${\sim}0.5~{\rm ppm}$ NV density, we estimate that ${\sim}2\times10^{10}$ NV centers are optically interrogated. Of these NV centers, ${\sim}1/4$ (${\sim}5\times10^{9}$) belong to the NV-axis sub-ensemble with $f_{\pm}$ spin resonances addressed by the microwaves.

An arbitrary waveform generator (Teledyne LeCroy T3AFG80) is used to generate AC test magnetic fields via an external loop surrounding the diamond sensor. The loop is formed from copper magnet wire (diameter $1.3~\rm mm$) that is wound three times in a circle of diameter $55~\rm mm$. In Fig.~\ref{fig5}(a), we used a modified loop that had two windings and an elliptical shape with ${\sim}55{\mbox{-}}\rm mm$-long major axis and ${\sim}20{\mbox{-}}\rm mm$-long minor axis.

A transistor-transistor logic (TTL) pulse card (SpinCore PBESR-PRO-500) is used to generate and synchronize pulse sequences and a data acquisition card (National Instruments USB-3631) is used to digitize the photodetector signal. 

\subsection{MW components}
The following microwave components were used (part numbers are for Mini-Circuits unless otherwise noted). 

When addressing the $f_-$ transition using MW carrier frequencies below $2.5~{\rm GHz}$ [Figs.~\ref{fig1}(d),\ref{fig2} (a-c)], we used ZHL-30W-252-S+ as the power amplifier. Otherwise, we used ZVE-3W-83+ followed by RF-Lambda RP02G06GSPA. 

For experiments presented in Fig.~\ref{fig2} (b,c), we used ZX05-42MH-S+ as the frequency mixer when both MW carrier frequencies were ${\lesssim}4~{\rm GHz}$ and ZMX-7GHR otherwise. In both cases, we used a combination of SHP-700+ and VLFX-450+ on the mixer output as an effective bandpass filter. To compensate for losses in the mixer and filters, we used an additional pre-amplifier before the power amplifier: ZX60-53LNB-S+ for MW carrier frequency ${\lesssim}5~{\rm GHz}$, and either ZVA-1S3WA-S+ or ZVA-183WA-S+ otherwise.

We conducted a number of control experiments to check that the pre-amplifiers, power amplifiers, mixers, and frequency filters did not add additional phase noise. We did not observe additional NV sensor excess noise when swapping out different components. For example, in Fig.~\ref{fig2}(b), there are two measured values for the $f_{\rm G1}=3.9\mbox{-}4.2~\rm GHz$ frequency range. One set of values corresponds to the use of the lower-frequency MW components, and the other used the higher-frequency components. No systematic difference in NV sensor excess noise is observed. A similar control is presented in Fig.~\ref{fig2}(c) for $f_{\rm G2,1}=4.1~\rm GHz$.

For the experiment presented in Fig.~\ref{fig5}(b), we used a diplexer Marki Microwave DPXN-0R5 to combine the microwave field and test signal.

In all experiments, the microwave power was adjusted to produce a $\pi$ pulse length of around $40-60~\rm ns$, depending on the carrier frequency. The normalized peak-to-peak amplitude of the processed photodetector signal during Rabi oscillations is typically ${\sim}6\%$.

\subsection{Gradiometry phase-noise cancellation}
\label{app:gradiometry}

\begin{figure}[htb]
    \includegraphics[width=0.99\columnwidth]{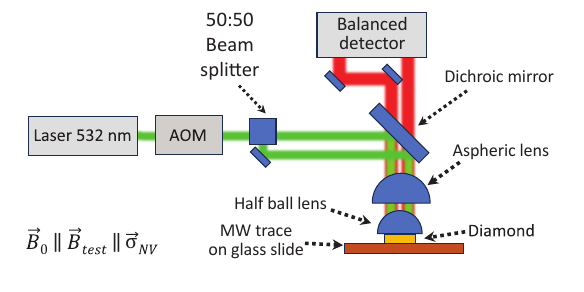}
    \caption{\textbf{Schematic of the gradiometry setup}. The primary difference from Fig.~\ref{fig1}(b) is that the excitation beam is split and focused to two spots on the diamond. The fluorescence from each spot is sent to different channels of the balanced photodetector. 
    }
  \label{setup_cancelation}
\end{figure}

For the two-point gradiometry experiments presented in Fig.~\ref{fig5}, we used a modified experimental setup, Fig.~\ref{setup_cancelation}. The excitation laser beam is split into two beams, which are focused to separate spots on the diamond. The fluorescence from each spot is collected and sent to different channels of the balanced photodetector (``ch 1'', ``ch 2''). The fluorescence spots are close enough together (${\sim}0.2~{\rm mm}$) that they experience approximately the same microwave field, meaning that both channels also have the same MW phase properties. By replacing the laser light with fluorescence from ch 2 on the balanced photodetector, we still compensate for laser intensity fluctuations, as both channels respond to the same laser intensity fluctuations. This scheme allows us to subtract the impact of MW phase noise by measuring the difference between the two channels of the balanced photodiode.

For these experiments, we used one of the out-of-diamond-plane NV axes. This was chosen because, in our setup, a test signal applied through the MW trace [Fig.~\ref{fig5}(b)] produces a field that is orthogonal to the in-plane NV axes. Producing a test field with a significant component along the NV axis was only possible for the two out-of-plane NV axes. To align $\vec{B_0}$ with an out-of-plane NV axis, we rotated the diamond in the electromagnet and modified the optical beam path accordingly.

\subsection{Magnetic field geometry}
\label{app:Bgeometry}
The bias magnetic field $\vec{B_0}$ is produced by an electromagnet, and the direction of $\vec{B_0}$ is aligned along one of the NV axes by rotating the diamond. For all measurements in Figs.~\ref{fig1}, \ref{fig2}, \ref{fig4}, and \ref{fig3}, $\vec{B_0}$ is aligned with an NV axis in the plane of the diamond faces. For two-point gradiometry, Fig.~\ref{fig5}, the magnetic field is aligned with an out-of-plane NV axis.

For almost all the measurements, the test field is also aligned along one of the NV axes. The exception is the experiment presented in Fig.~\ref{fig5}(b), where a gradient test field is applied through the MW trace. The gradient test field is approximately normal to the surface (at the location of the fluorescence spots) and thus makes a ${\sim}35\degree$ angle with respect to the relevant out-of-plane NV axis for our (110) diamond.

\subsection{Specifics of phase-noise spectroscopy measurements}
\label{app:specifics}

\begin{figure}[htb]
      \includegraphics[width=\columnwidth]{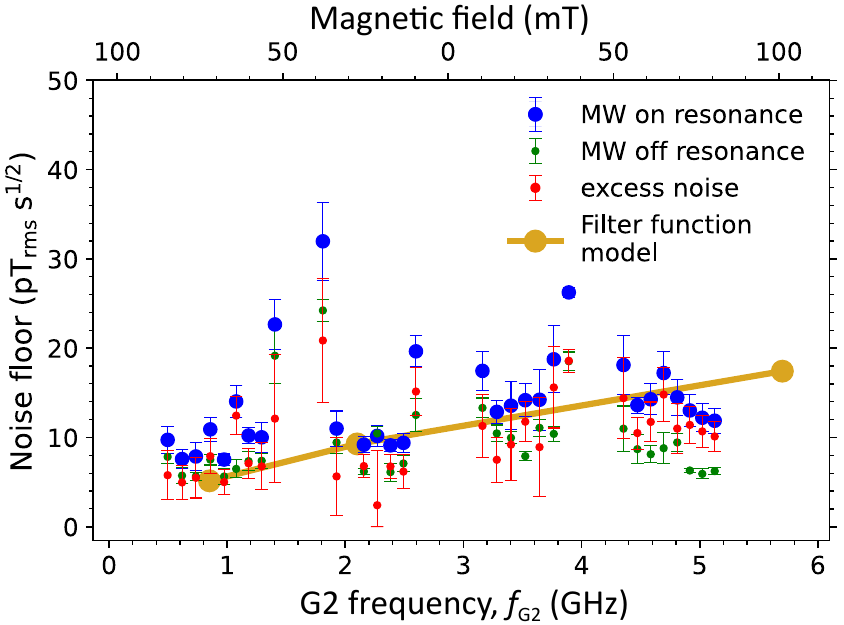}
    \caption{NV sensor noise floor for MW generator G2 tuned to a $f_{\pm}$ spin transition at different magnetic fields. For G2 carrier frequency $f_{\rm G2}<2.87~{\rm GHz}$, microwaves are tuned to the $f_-$ spin transition, and otherwise they are tuned to $f_+$.}
  \label{fig8}
\end{figure}

In Fig.~\ref{fig2}(a), we presented the NV sensor excess noise as a function of G1 MW carrier frequency. The $f_{\pm}$ resonances were swept by tuning $B_0$. Figure~\ref{fig8} shows a similar measurement using G2 under an XY8-6 pulse series. As in Fig.~\ref{fig2}(a), to generate this plot, $B_0$ is varied in the range $20\mbox{-}80~{\rm mT}$. For a given value of $B_0$, the noise floor is measured for $f_{\rm G2}$ tuned to a $f_{\pm}$ resonance, as well as the off-resonance case, and the excess noise $\eta_{\rm ex,\pm}$ is inferred. In Fig.~\ref{fig8}, the data are noisy, but there are hints of the same general trend of an increase in NV sensor noise with MW carrier frequency.

For the mixer experiments in Fig.~\ref{fig2}(b,c), we hold $B_0=80.7~\rm mT$ constant. A frequency mixer is used to produce a MW signal at the difference frequency between two inputs. For Fig.~\ref{fig2}(b), we use G1 and G2 as inputs, and for Fig.~\ref{fig2}(c) we use two channels of G2. The source generating the higher-frequency input is the one that is I/Q phase modulated in the XY8-$N_r$ pulse series. The difference frequency is always set to be $611~\rm MHz$. This configuration allows us to vary the MW generator frequency without having to vary $B_0$ or the NV transition frequencies. We can thus eliminate any effects due to the frequency response of different MW components, as well as the NV sensor's dependence on magnetic field (e.g. the vicinity of $B_0=0$, fields corresponding to NV cross-relaxations with other electron spins~\cite{LAZ2021}, and fields corresponding to $^{13}$C-induced NV echo envelope collapses \cite{MAZ2008}).

In Fig.~\ref{fig2}(c), the measurement point at the lowest frequency was obtained by using the mixer to create the frequency sum ($f_{\rm G2,1}+ f_{\rm G2,2}=310.5~{\rm MHz} + 300.0~{\rm MHz} = 610.5~{\rm MHz}$). The reason why we did this is because the lowest-frequency pair we were able to use to generate a clean $611~{\rm MHz}$ difference signal was approximately $\{0.3,0.911\}~{\rm GHz}$. The additional sum-frequency point allowed us to extend our observations to lower frequencies. Note that the phase noise of the mixer output is always assumed to be the quadrature sum of the phase noise of the individual inputs, see \ref{app:mixer}.

\subsection{Specifics of injected phase noise}
\label{app:noise_prep}

For the noise injection experiments of Figs.~\ref{fig4} and \ref{fig3}, voltage noise was generated using an arbitrary waveform generator and sent to the phase modulation port of G3. The random walk waveform has ${\sim}1.6{\times}10^6$ points and the waveform is repeated continuously in a loop. Each waveform is the concatenation of $1000$ random walk sequences. Each sequence has a variable length of mean $1600$ points and standard deviation $100$ points. The variable length is chosen to suppress any signatures of the sequence length in the frequency domain. The choice of a $1600$-point mean sequence length was made to ensure that voltage extrema in any sequence did not exceed the maximum input voltage of the phase modulation port. Each phase jump (the difference between adjacent points) is drawn from a normal distribution with a standard deviation $\sigma_{\rm rw}$ specified for each experiment. Varying $\sigma_{\rm rw}$ is realized by varying the G3 phase modulation port's scale factor. The first and last points of the waveform are set to 0 to accommodate the AWG's resetting behavior.

The white noise waveform comprises 1.6 million points drawn from a zero-mean normal distribution. The first and last points of the waveform are also set to 0, and the waveform repeats in a continuous loop.

The noise generated in this manner is not truly random. To verify that our injected noise protocol is still a good approximation to random-walk or white noise, we conducted experiments with different waveforms that were generated in the same manner. The results were consistent.

In all phase-noise-injection experiments, a $11~{\rm MHz}$ low-pass filter was used on the AWG output. The phase modulation port of G3 also has a pass band of DC to ${\sim}10~{\rm MHz}$. The limited bandwidth had some implications on the effective values of $\sigma_{\rm rw}$ and $\sigma_{\rm wh}$. We independently measured the frequency response by mixing two channels of G3, one of which was phase modulated with the noise waveforms. From analysis of the homodyne signal, we were able to directly measure $\sigma_{\rm wh}$. We found that this value was ${\sim}20\%$ lower than the expected value if the noise injection had infinite bandwidth. In all measurements in the paper, we report the slightly lower value that was directly measured. For $\sigma_{\rm rw}$, we find the limited bandwidth also leads to attenuation, but the exact value depends on the sampling interval. However, for most experiments in the text, the sampling interval is consistent with a ${\sim}15\%$ attenuation, so we use this value throughout.

\section{Calibrated test fields}
\label{app:testcal}
To calibrate the AC test signals, we used a method similar to that of Refs.~\cite{GLE2018,SIL2023}. A voltage signal, with amplitude $V_{\rm test}$ and frequency $f_{\rm test}$, is applied to the test loop, producing a homogenous AC test field within the diamond. The amplitude $V_{\rm test}$ is varied and the NV signal is recorded at each value of $V_{\rm test}$. The magnitude of the NV fluorescence photodetector signal due to the test field, $V_{\rm nv}$, is given by:
\begin{equation}
\label{eq:NVsat}
V_{\rm nv}=V_{\rm max}|\sin(4\sqrt{2}\,\kappa\, V_{\rm test}\gamma_{\rm nv}\tau_{\rm tot})|,
\end{equation}
where $V_{\rm max}$ is the maximum NV signal magnitude, $\kappa=B_{\rm test}/V_{\rm test}$ is a fitted scaling factor, and $\tau_{\rm tot}=8(2\tau+t_{\rm \pi})N_{\rm r}$ is the interval between $\pi/2$ pulses in the XY8-$N_r$ sequence (see Sec.~\ref{sec:exptsetup} of the main text). The fitted calibration factor, $\kappa$, allows us to convert the applied test signal amplitude to magnetic field units. 

\begin{figure}[htb]     
\includegraphics[width=\columnwidth]{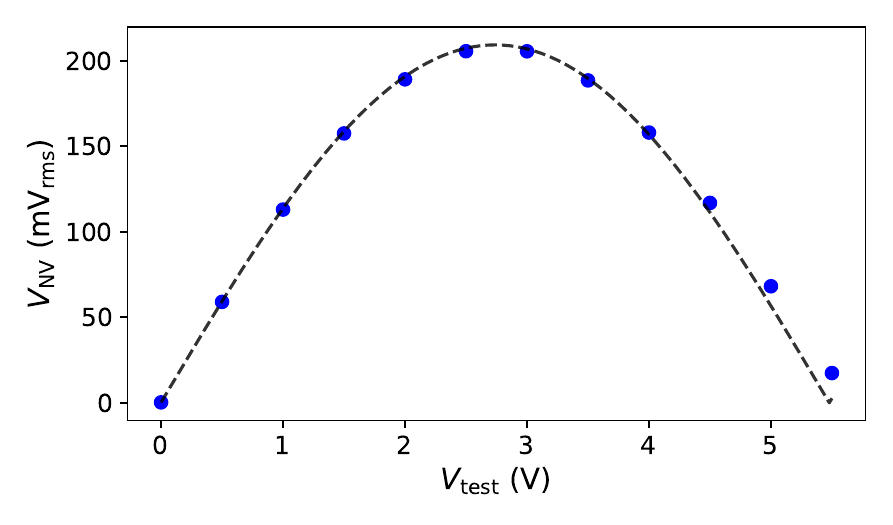}
    \caption{Plot of the NV sensor test-signal peak $V_{\rm nv}$ as a function of the voltage amplitude of the test signal, $V_{\rm test}$. A fit to Eq.~\ref{eq:NVsat} reveals $\kappa$, the voltage-to-field conversion factor.}
  \label{Saturation}
\end{figure}

Figure~\ref{Saturation} shows an example plot of $V_{\rm nv}(V_{\rm test})$ along with a fit to Eq.~\eqref{eq:NVsat} used to calibrate the test signal in Fig.~\ref{fig1}(d). We typically repeat the calibration process before and/or after acquiring a data set to account for drifts of the test loop position or NV collection region.

\section{Independent MW phase-noise measurements}
\label{appendix:Direct_PN}

The phase noise spectra, $\mathcal{L}(f)$, of the G1 and G3 MW generators were measured directly with a Berkeley Nucleonics 7000 Phase Noise Tester. Further direct G1 phase noise spectra were acquired with a Rohde \& Schwarz FSWP26 [see Fig.~\ref{fig2}(d)]. The G1 $\mathcal{L}(f)$ curves obtained with the two instruments were similar and also comparable to the manufacturer's specifications. We did not measure $\mathcal{L}(f)$ curves of G2 directly. However, we found that the directly-measured $\mathcal{L}(f)$ curves for G3 (a similar model from the same manufacturer) matched closely those reported as ``typical'' by the manufacturer. This lends confidence that our use of the manufacturer-specified $\mathcal{L}(f)$ curves for G2 is a decent approximation.

All direct $\mathcal{L}(f)$ measurements were performed in a continuous-wave mode, without pulsing, but with I/Q modulation enabled. When the MW power input to the amplifier was low ($\lesssim-10~{\rm dBm}$), the single-sideband (SSB) phase noise values were higher than specifications at higher offset frequencies ($\gg1~{\rm kHz}$); this effect is negligible at the high input powers used in our experiments for the frequency bands of interest. When adding a MW switch, and/or using a power amplifier and attenuators, there was no significant effect on the measured $\mathcal{L}(f)$ curves aside from minor variations at offset frequencies $>10~{\rm MHz}$.

\section{Data acquisition and processing} 
\label{appendix:processing}

NV sensor noise signals were acquired in two modes. In the first mode, used for most experiments including Figures~\ref{fig2}(a-c,f), \ref{fig4} and \ref{fig3}, a synchronized XY8-$N_r$ pulse series of length ${\sim}30~{\rm ms}$ is applied, and the test signal is synchronized with the start of the series using the burst mode of a function generator. We repeat the pulse series $N_{\rm avg}\approx35$ times, and the photodetector time trace of each series is averaged in the time domain. The absolute value of the Fourier transform of the averaged data is computed, and the noise floor is extracted as described in Sec.~\ref{app:noise}. The above process takes ${\sim}1~{\rm s}$, and it is repeated for ${\gtrsim}5$ runs to gather statistics. The median and standard deviation of the noise floors is computed, serving as, respectively, the point estimate and uncertainty in the figures.

The second mode is used for data presented in Figs.~\ref{fig1}(d) and~\ref{fig5}. Here, a synchronized XY8-$N_r$ pulse series is applied continuously for ${\sim}150~{\rm s}$. The photodetector time trace is split into $1{\mbox{-}}{\rm s}$ intervals, and the absolute-value-squared of the Fourier transform (power spectral density) of each interval is computed. The square root of the mean of the power spectral density is obtained, producing a final ``smoothed'' spectrum. In Sec.~\ref{sec:exptsetup} of the main text, we call this spectrum ``the root-mean-squared average~\cite{GRA2024} of the absolute value of the Fourier
transform of each interval.''

\begin{table}[htb]
    	\centering
    	\begin{tabular}[b]{c | c | c | c | c | c | c }
		Fig. & $\,N_r\,$ & $\,t_{\pi}\,{\rm (ns)}\,$ & $\,\tau\,{\rm (ns)}\,$ & $\,\tau_{\rm tot}\,({\rm \upmu s})\,$ & $\,t_{\rm dead}\,({\rm \upmu s})\,$ & $\,f_{\rm xy8}\,({\rm kHz})\,$\\\hline
        \ref{fig1}(d) & 8 & 48 & 522 & 69.9 & 15.0 & 458\\  
		\ref{fig2}(a) & 6 & 48 & 622 & 62.0 & 14.9 & 387\\
		\ref{fig2}b & 8 & 48 & 522 & 69.9 & 14.7 & 458\\
		\ref{fig2}c & 8 & 48 & 522 & 69.9 & 14.7 & 458\\
		\ref{fig2}f & 1 & 58 & 3844 & 62.0 & 27.5 & 64.5\\
		$''$ & 2 & 58 & 2046 & 66.4 & 23.1 & 120\\
		$''$ & 3 & 58 & 1388 & 68.1 & 21.4 & 176\\
		$''$ & 4 & 58 & 1100 & 72.3 & 17.1 & 221\\
		$''$ & 5 & 58 & 872 & 72.2 & 17.2 & 277\\
		$''$ & 6 & 58 & 720 & 72.0 & 17.3 & 333\\
		$''$ & 7 & 58 & 612 & 71.9 & 17.4 & 389\\
		$''$ & 8 & 58 & 532 & 71.9 & 17.5 & 445\\
		$''$ & 9 & 58 & 468 & 71.7 & 17.5 & 502\\
		$''$ & 10 & 58 & 418 & 71.7 & 17.5 & 558\\
		$''$ & 11 & 58 & 378 & 71.8 & 17.6 & 613\\
		$''$ & 12 & 58 & 344 & 71.8 & 17.6 & 668\\
		$''$ & 16 & 58 & 250 & 71.7 & 17.7 & 893\\
		$''$ & 17 & 58 & 232 & 71.3 & 17.6 & 954\\
		\ref{fig5}(a-c) & 7 & 46 & 612 & 71.1 & 17.1 & 394\\
		\end{tabular}
    \caption{Experimental parameters used in the measurements presented in Figs.~\ref{fig1}, \ref{fig2}, and \ref{fig5}.}
    \label{tbl:fig1_2_5data}
\end{table}

\begin{table}[hbt]
	\centering
	\begin{tabular}[b]{c | c | c | c | c | c | c }
		Fig. & $N_r$ & $\,t_{\pi}\,{\rm (ns)}\,$ & $\,\tau\,{\rm (ns)}\,$ & $\,\tau_{\rm tot}\,({\rm \upmu s})\,$ & $t_{\rm dead}\,({\rm \upmu s})$ & $f_{\rm xy8}\,({\rm kHz})$\\\hline
		\ref{fig3}(a) & 8 & 48 & 522 & 69.9 & 14.7 & 458\\
		\ref{fig3}(b),\ref{fig4}(a) & 8 & 48 & 522 & 69.9 & 14.7 & 458\\
		\ref{fig3}(c),\ref{fig4}(b) & 1 & 44 & 3768 & 60.6 & 27.7 & 66.0\\
		$''$ & 2 & 44 & 2015 & 65.2 & 23.0 & 124\\
		$''$ & 3 & 44 & 1372 & 66.9 & 21.3 & 179\\
		$''$ & 4 & 44 & 1090 & 71.2 & 17.0 & 225\\
		$''$ & 5 & 44 & 866 & 71.0 & 17.1 & 282\\
		$''$ & 6 & 44 & 718 & 71.0 & 17.2 & 338\\
		$''$ & 7 & 44 & 610 & 70.8 & 17.3 & 396\\
		$''$ & 8 & 44 & 532 & 70.9 & 17.3 & 451\\
		$''$ & 9 & 44 & 470 & 70.9 & 17.3 & 508\\
		$''$ & 10 & 44 & 420 & 70.7 & 17.4 & 566\\
		$''$ & 11 & 44 & 380 & 70.8 & 17.4 & 622\\
		$''$ & 12 & 44 & 346 & 70.7 & 17.4 & 679\\
		$''$ & 13 & 44 & 318 & 70.7 & 17.4 & 735\\
		$''$ & 16 & 44 & 254 & 70.7 & 17.5 & 906\\
		$''$ & 17 & 44 & 238 & 70.7 & 17.5 & 962\\
		\ref{fig3}(d),\ref{fig4}(c) & 1 & 44 & 610 & 10.1 & 17.6 & 396\\
		$''$ & 2 & 44 & 610 & 20.2 & 17.5 & 396\\
		$''$ & 3 & 44 & 610 & 30.3 & 17.5 & 396\\
		$''$ & 4 & 44 & 610 & 40.5 & 17.4 & 396\\
		$''$ & 5 & 44 & 610 & 50.6 & 17.4 & 396\\
		$''$ & 6 & 44 & 610 & 60.7 & 17.3 & 396\\
		$''$ & 7 & 44 & 610 & 70.8 & 17.3 & 396\\
		$''$ & 8 & 44 & 610 & 80.9 & 17.2 & 396\\
		$''$ & 9 & 44 & 610 & 91.0 & 17.2 & 396\\
		$''$ & 10 & 44 & 610 & 101.1 & 17.1 & 396\\
		$''$ & 11 & 44 & 610 & 111.2 & 17.1 & 396\\
		$''$ & 12 & 44 & 610 & 121.3 & 17.0 & 396\\
		$''$ & 13 & 44 & 610 & 131.5 & 17.0 & 396\\
		$''$ & 14 & 44 & 610 & 141.6 & 16.9 & 396\\
		$''$ & 15 & 44 & 610 & 151.7 & 16.9 & 396\\
		$''$ & 16 & 44 & 610 & 161.8 & 19.3 & 396\\
		\end{tabular}
	\caption{Experimental parameters used for measurements presented in Figs.~\ref{fig3} and \ref{fig4}.}
	\label{tbl:fig3data}
\end{table}

Experimental parameters used to generate Figs.~\ref{fig1}, \ref{fig2}, and \ref{fig5} are shown in Table~\ref{tbl:fig1_2_5data}. Experimental parameters used to generate Figs.~\ref{fig4} and \ref{fig3} are shown in Table~\ref{tbl:fig3data}. Some parameters used in the analytic calculations in these figures are the median of experimental values, as indicated in their insets.

\subsection{Noise floor estimation}
\label{app:noise}
The NV sensor noise floor is computed as the median of the absolute value of the Fourier transform in ``spike-free'' regions. The noise floor can be verified visually, as seen in Fig.~\ref{fig1}(d). For experiments requiring the estimation of multiple noise floors, the process of defining spike-free regions is automated. First, the frequency range of $0\mbox{-}1~{\rm kHz}$ and a ${\sim}1~{\rm kHz}$ band about the test signal frequency are excluded. Next, the median and standard deviation of the remaining spectrum, excluding the highest $10\%$ of spectral points, is computed. Spectral points that are more than four standard deviations larger than the median are identified as spikes and excluded. The noise floor is then given by the median of the remaining spectrum.

This automated process to calculate the noise floor was tuned and routinely checked for accuracy based on visual inspection of the spectrum. In some datasets, this procedure is replaced by visually locating a spike-free region near the test signal frequency and computing the median of the noise floor in this region.

\section{Photoelectron-shot-noise-limited sensitivity}
\label{appendix:psn}
The minimum detectable magnetic field due to photoelectron shot noise for a single XY8-$N_r$ readout is given by~\cite{TAY2008,SIL2023}:
\begin{equation}
\label{eq:BminSingle2}
\Delta B_{\rm single}\approx\frac{\xi}{4\,\gamma_{\rm nv}\,C\,\tau_{\rm tot}\sqrt{N_{\rm ph}}}.
\end{equation}
Here $\xi=\sqrt{2\,(1+t_{\rm r}/t_{\rm n})}$ accounts for the extra photoelectron noise due to the balanced detection and normalization procedure, where $t_{\rm r}$ is the readout duration at the start of the laser pulse, $t_{\rm n}$ is the duration of normalization at the end of the laser pulse, $C$ is the effective contrast of the XY8-$N_r$ sequence, and $N_{\rm ph}$ is the number of photoelectrons detected in the readout phase.

The theoretical limit for the magnetometer sensitivity due to photoelectron shot noise is defined as:
\begin{equation}
\label{eq:etaminSingle2}
\eta_{\rm psn}=\Delta B_{\rm single} \sqrt{\tau_{\rm tot}+t_{\rm dead}}\approx\frac{\xi}{\sqrt{\delta}}\frac{1}{4\gamma_{\rm nv}C\sqrt{\tau_{\rm tot}\,N_{\rm ph}}},
\vspace{3 mm}
\end{equation}
where $\delta{=}\tau_{\rm tot}/(\tau_{\rm tot}{+}t_{\rm dead})$ is the measurement duty cycle. Using the experimental parameters of Fig.~\ref{fig1}(d) ($t_{\rm n}{=}4~{\rm \upmu s}$, $t_{\rm r}{=}1.5~{\rm \upmu s}$, $\tau_{\rm tot}{=}69.9~{\rm \upmu s}$, $t_{\rm dead}{=}15~{\rm \upmu s}$, $C{=}0.013$, $N_{\rm ph}{=}1.23{\times}10^9$), we estimate $\eta_{\rm psn}\,{\approx}\,4.3~{\rm pT_{rms}\,s^{1/2}}$. To convert to a noise floor in the absolute value of the Fourier transform, this quantity is multiplied by a factor $\sqrt{\pi/2}\approx1.253$, resulting in $\eta_{\rm psn}\approx5.4~{\rm pT_{rms}\,s^{1/2}}$. The latter is the value that is directly comparable to our experimental noise-floor measurements. 

In Eqs.~\eqref{eq:BminSingle2} and \eqref{eq:etaminSingle2}, the estimation of $N_{\rm ph}$ is made by recording the peak voltage of the photodetector fluorescence-channel's monitor port. While this level varied somewhat from experiment to experiment, for the data in Fig.~\ref{fig1}(d) it was $V_{\rm mon}\approx0.39~{\rm V}$ at $50~\ohm$ impedance. Taking into account the difference in gain between the two ports inferred from the manufacturer specifications, this is equivalent to an effective voltage in the RF difference port (at $50~\ohm$ impedance) of $V_0\approx23~{\rm V}$. We could not use the RF port to measure $V_0$ directly because it saturates at ${\sim}3.5~{\rm V}$. The estimation of $C$ is obtained by taking the peak value $V_{\rm max}=0.2~{\rm V_{rms}}$ in Fig.~\ref{Saturation}. From Eq.~\eqref{eq:NVsat} (see also Ref.~\cite{SIL2023}), the contrast is given by $C\approx V_{\rm max}\sqrt{2}/V_0\approx0.013$.

\section{Early attempts to understand the excess noise}
\label{appendix:earlytries}
As mentioned in the main text, when we first observed the presence of excess noise, $\eta_{\rm ex,\pm}$, we did several tests to confirm its nature. Things that did not change the noise behavior included:

\begin{enumerate}
    \item Swapping amplifiers. Instead of an RF-Lambda RP02G06GSPA, we used a Mini-Circuits ZHL-25W-63+. A small increase in the contrast, $C$ of the magnetometer was observed, but $\eta_{\rm ex, \pm}$ was unchanged. We believe the improvement in $C$ is due to the higher output power of the Mini-Circuits amplifier improving the microwave pulse fidelity.
    \item Modified our method of I/Q modulation. Instead of using switches to apply DC voltages to the I/Q modulation ports of the signal generators, we used a single switch toggling between two out-of-phase 300 MHz signals from a function generator. No perceptible change in noise floor was observed.
    \item Reduced the MW Rabi frequency, increasing the pulse lengths. We observed a small reduction in $C$, but $\eta_{\rm ex, \pm}$ was unchanged. Further reduction in MW Rabi frequency degraded the sensitivity to the point where the $\eta_{\rm ex, \pm}$ could no longer be distinguished from the photoelectron shot-noise. 
    \item We changed the lengths of the cables carrying TTL signals to the I/Q modulation ports of the signal generators. Cable lengths were adjusted such that the phase toggling for the MW pulses happened exactly between any two pulses firing. We observed a small improvement in $C$, but $\eta_{\rm ex, \pm}$ was unchanged.
\end{enumerate}

\section{Phase noise when mixing two frequencies}
\label{app:mixer}

As discussed in Sec.~\ref{sec:freqGendependece} of the main text, a mixer was used to probe the impact of phase noise on the NV sensor noise floor as a function of microwave carrier frequency. For pedagogical reasons, we derive the well-known operation of an ideal mixer here. 

If a mixer is assumed to be perfect, two input signals $v_{1}(t)=v_{01} \sin\left(\omega_1 t + \alpha_1\right)$ and $v_2(t)=v_{02}\sin\left(\omega_2 t + \alpha_2\right)$ in the input would yield an output signal:
\begin{equation}
v_3=v_{03}\sin\left(\omega_1 t + \alpha_1\right)\sin\left(\omega_2 t + \alpha_2\right).
\end{equation}
By applying the product-to-sum identity the above expression can be rewritten as
\begin{equation}
\label{eq:diffsum}
\begin{split}
    v_3=\frac{v_{03}}{2}\{ \cos[(\omega_2-\omega_1)t + (\alpha_2 - \alpha_1)]\,+ \\  \cos[(\omega_2+\omega_1)t  +
    (\alpha_2+\alpha_1)]\}.
\end{split}
\end{equation}

In the experiments depicted in Fig.~\ref{fig2}(b,c) a filter is used to keep only the difference-frequency term. The exception is the lowest-frequency point of Fig.~\ref{fig2}(c), where only the sum-frequency term is kept. In either case, applying the filtering process to Eq.~\eqref{eq:diffsum} leaves:
\begin{equation}
\Tilde{v}_3 = \frac{v_{03}}{2}\cos(\Delta\omega t + \Delta \alpha),
\end{equation}
where $\Delta \omega = \omega_2\pm\omega_1$ and $\Delta \alpha = \alpha_2\pm\alpha_1$. 

In the presence of phase noise, the $\alpha_1$ and $\alpha_2$ phases of the signals fluctuate in time. If the two signals are generated independently, then $\alpha_1$ and $\alpha_2$ are uncorrelated, and the standard deviation of the output phase, $\Delta \alpha$, is:
\begin{equation}
\label{eq:mix_phase}
\sigma_{\Delta \alpha}=\sqrt{\sigma^2_{\alpha_1}+\sigma^2_{\alpha_2}},
\end{equation}
where $\sigma_{\alpha_1}$ and $\sigma_{\alpha_2}$ are the standard deviations of the input phases, $\alpha_1$ and $\alpha_1$, respectively. 

From equation~\ref{eq:mix_phase} it can be seen that if $\sigma_{\alpha_1}\gg\sigma_{\alpha_2}$ then $\sigma_{\Delta \alpha}\approx \sigma_{\alpha_1}$. This is the approximation applied for the measurements in Fig.~\ref{fig1}(b), where the phase noise of G1 is much higher than that of G2. If $\sigma_{\alpha_1}\approx \sigma_{\alpha_2}$ then $\sigma_{\Delta\alpha}\approx\sqrt{2}\sigma_{\alpha_1}$. This is the approximation applied for the measurements in Fig.~\ref{fig2}(c), where two independent channels of the same generator, G2, are used. The assumption that $\sigma_{\alpha_1}\approx \sigma_{\alpha_2}$ starts to break down if the ratio of the carrier frequencies $\omega_1/\omega_2$ deviates substantially from unity. In Fig.~\ref{fig2}(c), this is the region $f_{\rm G2,1}\lesssim1~{\rm GHz}$ where we did not acquire data using the difference frequency. However, the approximation still holds when using the sum frequency, as long as the two carrier frequencies are comparable, as was the case for our lowest-frequency measurement in Fig.~\ref{fig2}(c). We thus were able to apply the approximation uniformly to all the data in Fig.~\ref{fig2}(c).

\section{Frequency-domain model using filter-function approach}
\label{appendix:filterfunc}
The phase-noise-limited equivalent magnetic sensitivity can be calculated for arbitrary phase noise character, as long as the MW phase-noise spectrum is known. For this, we use the independently-measured phase noise spectrum of the MW signal generator (see Fig.~\ref{fig2}d) and the frequency-domain filter function of the pulse sequence (identical for CPMG or XY8-$N_{\rm r}$). The single-sideband power spectral density of the local oscillator's phase fluctuations in a MW signal generator, $S_{\phi}(f)$ (in units of ${\rm rad^{2}/Hz}$), can be written as~\cite{RUB2008}:
\begin{equation}
\label{eq:MWgenPSD}S_{\phi}(f)\,{=}\,2\times10^{\frac{\mathcal{L}(f)}{10}},
\end{equation}
where $\mathcal{L}(f)$ is the single-sideband power spectral density of the phase noise expressed in logarithmic units of dBc/Hz, and $f$ is the carrier offset frequency.

The standard deviation of the NV spin state's phase displacement ($\sigma_{\phi}$) due to MW phase errors is given by:
\begin{equation}
\label{eq:NVphasSDGen}
\sigma_{\phi}^{2}=\int_{0}^{\infty} S_{\phi}(f)\,\mathcal{F}(f)\,df\approx\int_{0}^{f_c} S_{\phi}(f)\,\mathcal{F}(f)\,df,
\vspace{3 mm}
\end{equation}
where $\mathcal{F}(f)$ is the frequency-domain filter function of the measurement sequence and $f_c$ is a cutoff frequency above which $S_{\phi}(f)\mathcal{F}(f)\approx0$. In practice, the value of $f_c$ is influenced by low-pass filtering of MW phase noise due to the MW circuit spectral response, any bandpass filters in the MW chain, and dissipation from the various MW components and coaxial cables (\ref{app:cutoff}). The filter function for a pulse sequence with equally-spaced $\pi$ pulses (including CPMG and XY8-$N_r$) is given by~\cite{Bie2011}:
\begin{equation}
\begin{split}
\label{eq:FiltrFuncXY8}\mathcal{F}(f)=|1+&(-1)^{N+1} e^{i 2 \pi f\tau_{\rm tot}}\,{+}\\ &~~~ 2 \sum_{j=1}^{N}(-1)^{j} e^{i\,\frac{j-1/2}{N}\,2 \pi f \tau_{\rm tot}}\,G(t_{\pi})|^{2},
\end{split}
\end{equation}
where $G(t_{\pi})$ is a term that modifies the filter function to take into account the finite-length of the $\pi$ pulses, $t_{\pi}$. In the limit $t_{\pi}\rightarrow0$, the NV total phase accumulation time is $\tau_{\rm tot}\,{=}\,2\,N\,\tau$ and $G(t_{\pi})\,{=}\,1$. If $t_{\pi}$ can not be neglected, we take $\tau_{\rm tot}\,{=}\,N\,(2\,\tau\,{+}\,t_{\pi)}$ and $G(t_{\pi})\,{=}\, \cos{(\pi f t_{\pi})}$~\cite{Bie2011}.

Fluctuations in the phase of the NV spin state, Eq.~\eqref{eq:NVphasSDGen}, result in an equivalent magnetic noise. For a single readout of a pulse sequence, the equivalent magnetic noise standard deviation is given as~\cite{TAY2008}:
\begin{equation}
\label{eq:MagnticNois} \Delta B_{\rm single}\,\approx\,\frac{\sigma_{\rm \phi}}{4\,\gamma_{\rm nv}\,\tau_{\rm tot}},
\end{equation}
where $\gamma_{\rm nv}$ is the NV gyromagnetic ratio and $\tau_{\rm tot}$ is the NV total phase accumulation (or sensing) time. When averaging multiple pulse sequences, each of duration $\tau_{\rm tot}+t_{\rm dead}$, for a total acquisition time $T$, there are $T/(\tau_{\rm tot}+t_{\rm dead})$ independent measurements. In this case, the equivalent magnetic noise standard deviation becomes:
\begin{equation}
\label{eq:MagnticNoisTot} \Delta B_{\phi}\,=\,\frac{\Delta B_{\rm single}}{\sqrt{T/(\tau_{\rm tot}+t_{\rm dead})}}.
\end{equation}
Using Eqs.~\eqref{eq:MagnticNois} and \eqref{eq:MagnticNoisTot}, the MW-phase-noise limited equivalent magnetic sensitivity is given by:
\begin{equation}
\label{eq:SenMWPhasNois} \eta_{\phi}~{=}~\Delta B_{\phi}\,\sqrt{T}\,{\approx}\,\frac{\sigma_{\rm \phi}}{4\,\gamma_{\rm nv}\sqrt{\tau_{\rm tot}}}\,\sqrt{1+t_{\rm dead}/\tau_{\rm tot}}.
\end{equation}

The NV sensor noise floor calculations shown in Fig.~\ref{fig2} were computed using Eqs.~\eqref{eq:SenMWPhasNois},~\eqref{eq:NVphasSDGen}, and~\eqref{eq:FiltrFuncXY8}. The parameters used in the calculations are shown in Table~\ref{tbl:NVMagntMWnoisCal}. The resulting calculated values of $\eta_{\rm ex}$ are listed in Tables~\ref{tab:SRS_calcvals} and~\ref{tab:SMU_calcvals}, using two approximations for $G(t_{\pi})$ in Eq.~\eqref{eq:FiltrFuncXY8}. In the first, we assume infinitesimal pulses $t_{\pi}\,{=}\,0$ and thus $G=1$. In the second, we assume finite pulses ($t_{\pi}\,{=}\,48\,{\rm ns}$) and $G(t_{\pi})=\cos{(\pi f t_{\pi})}$. We use the finite-pulse approximation for values in Fig.~\ref{fig2} of the main text, but the differences are small. Note that the magnetic noise floor of the absolute value of a Fourier transform ($\eta_{\rm ex}$) is defined slightly differently from the equivalent magnetic sensitivity ($\eta_{\phi}$) in Eq.~\eqref{eq:SenMWPhasNois}. To convert, we take $\eta_{\rm ex}=\eta_{\phi}\sqrt{\pi/2}\approx1.253\,\eta_{\phi}$.

\begin{table}[htb]
\centering
	\begin{tabular}[b]{r | c | c | c | c | c | c}
		Fig. & $N_{\rm r}$ & ~$t_{\pi}$ ($\rm ns$) & ~$\tau$ ($\rm ns$) & ~$\tau_{\rm tot}$ ($\rm \mu s$) & ~$t_{\rm dead}$ ($\rm \mu s$) & ~$f_{\rm xy8}$ ($\rm kHz$)\\
		\hline
		\ref{fig2}(a) &     6 &    48 &   622 &  62.0 &  14.85 &  390\\
		\ref{fig2}(b) &     8 &    48 &   522 &  69.9 &  14.85 &  461\\
		\ref{fig2}(c) &     8 &    48 &   522 &  69.9 &  14.85 &  461\\
		\ref{fig2}(f) &     - &    60 &   - &  71.9 &    17.1  &  -  \\
	\end{tabular}
	\caption{Parameters used to calculate $\eta_{\rm \phi}$, using Eqs.~\eqref{eq:NVphasSDGen}, \eqref{eq:FiltrFuncXY8}, and \eqref{eq:MagnticNoisTot}. The resulting $\eta_{\rm \phi}$ values are shown in Table \ref{tab:SRS_calcvals} and \ref{tab:SMU_calcvals} and plotted in Fig.~\ref{fig2}.
 }
	\label{tbl:NVMagntMWnoisCal}
\end{table}

\begin{table}[ht]
\begin{tabular}{c|c|c|c}
~Fig.~ & ~$f_{\rm G1}~{\rm (GHz)}$~ & ~$t_{\pi}~{\rm (ns)}$~ & ~$\eta_{\phi}~({\rm pT_{rms}\,s^{1/2}})$~\\ [0.5ex]
\hline
\ref{fig2}(a) & 0.61$^{\ast}$ & 0 & 24.9 \\
" & 0.61$^{\ast}$ & 48 & 23.0 \\
" & 1.00 (spec) & 0 & 30.2 \\
" & 1.00 (spec) & 48 & 27.5 \\
" & 1.00$^{\ast}$ & 48 & 28.2 \\
" & 2.50$^{\ast}$ & 0 & 56.0 \\
" & 2.50$^{\ast}$ & 48 & 54.1 \\
" & 6.00 (spec) & 0 & 107.0 \\
" & 6.00 (spec) & 48 & 104.9 \\
" & 6.00$^{\ast}$ & 48 & 123.0 \\
\ref{fig2}(b) & 0.61$^{\ast}$ & 0 & 26.7 \\
" & 0.61$^{\ast}$ & 48 & 24.7 \\
" & 1.00 (spec) & 0 & 32.5 \\
" & 1.00 (spec) & 48 & 29.7 \\
" & 1.00$^{\ast}$ & 0 & 32.4 \\
" & 1.00$^{\ast}$ & 48 & 30.2 \\
" & 2.50$^{\ast}$ & 0 & 60.1 \\
" & 2.50$^{\ast}$ & 48 & 58.0 \\
" & 6.00 (spec) & 0 & 113.5 \\
" & 6.00 (spec) & 48 & 111.2 \\
" & 6.00$^{\ast}$ & 48 & 129.9 \\
\end{tabular}
\caption{Calculated phase-noise-limited equivalent magnetic sensitivity values for G1, using the filter function model, as presented in Fig.~\ref{fig2}(a,b). * denotes $\mathcal{L}(f)$ was measured independently using a phase noise analyzer. ``spec'' denotes $\mathcal{L}(f)$ was extracted from the manufacturer's specifications sheet~\cite{M_SRS386}, shown for comparison purposes. The values of $\eta_{\rm ex}$ shown in Fig.~\ref{fig2} used finite pulse duration ($t_{\pi}=48~{\rm ns}$) and the measured $\mathcal{L}(f)$ spectra. In Fig.~\ref{fig2}(a,b), each of the $\eta_{\phi}$ values in this table are multiplied by a factor of $\sqrt{\pi/2}=1.253$ to compare to the experimental noise floor, $\eta_{\rm ex}$. Here, the maximum frequency offset is taken to be $f_c=0.1~{\rm GHz}$.}
\label{tab:SRS_calcvals}
\end{table}

\begin{table}[htb]
\begin{tabular}{c|c|c|c}
~Fig.~ & ~$f_{\rm G2}~{\rm (GHz)}$~ & ~$t_{\pi}~{\rm (ns)}$~ & ~$\eta_{\phi}~({\rm pT_{rms}\,s^{1/2}})$~\\ [0.5ex]
\hline
\ref{fig2}(c) & 0.85 & 0 & 5.4 \\
" & 0.85 & 48 & 4.1 \\
" & 2.10 & 0 & 9.4 \\
" & 2.10 & 48 & 7.4 \\
" & 5.70 & 0 & 15.9 \\
" & 5.70 & 48 & 13.9 \\
\end{tabular}
\caption{Calculated phase-noise-limited sensitivity values values for G2, using the filter function model, as presented in Fig.~\ref{fig2}(c). In all cases, $\mathcal{L}(f)$ was extracted from the manufacturer's specifications sheet~\cite{M_SMATE200A}. The values of $\eta_{\rm ex}$ shown in Fig.~\ref{fig2}(c) used finite pulse duration ($t_{\pi}=48~{\rm ns}$). In Fig.~\ref{fig2}(c), each of the $\eta_{\phi}$ values in this table are first multiplied by a factor of $\sqrt{2}$ to account for the mixer addition (\ref{app:mixer}) and then multiplied by a factor of $\sqrt{\pi/2}=1.253$ to compare to the experimental noise floor, $\eta_{\rm ex}$. Here, the maximum frequency offset is taken to be $f_c=0.1~{\rm GHz}$.}
\label{tab:SMU_calcvals}
\end{table}

\subsection{Oscillator Johnson phase noise limit: pulsed}
\label{app:Johnsonpulse}

A fundamental source of phase noise in any practical MW generator is that due to Johnson noise, which is frequency-independent (white) and only depends on temperature. If the local oscillator of a signal generator is driven by $1\,{\rm mW}$ ($0\,{\rm dBm}$) power near room temperature, the minimum MW phase noise of the output carrier in $1\,{\rm Hz}$ bandwidth is $\mathcal{L}(f)=\mathcal{L}_{\rm min}\approx-177\,{\rm dBc/Hz}$~\cite{HAT2003,RUB2008}. 

The variance in the NV spin's phase displacement, Eq.~\eqref{eq:NVphasSDGen}, due to the MW Johnson phase errors can be written as:
\begin{equation}
\label{eq:NVphasSDGenJohn}
\sigma_{\phi,J}^{2}\approx2\times 10^{\frac{\mathcal{L}_{\rm min}}{10}}\int_{0}^{f_c}\mathcal{F}(f)\,df.
\end{equation}
For the multipulse filter function in Eq.~\eqref{eq:FiltrFuncXY8}, the integral on the right-hand side of Eq.~\eqref{eq:NVphasSDGenJohn} is independent of the pulse length ($\tau_{\rm \pi}$) and spacing between the pulses ($\tau$) if $f_c\gtrsim 1/(2\,\tau_{\rm \pi})$, and it can be approximated as~\cite{BAL2016}:
\begin{equation}
\label{eq:FiltrIntgrat}
\int_{0}^{f_c}\mathcal{F}(f)\,df\,\approx
\begin{cases}
\,\,(4N+2)\times 2\,\pi\,f_c &  
  \tau_{\rm \pi}\rightarrow 0 \\
  \\
\,\,(2N+2)\times 2\,\pi\,f_c &  
  \tau_{\rm \pi}\neq 0. \\
\end{cases}
\vspace{3 mm}
\end{equation}
Using Eqs.~\eqref{eq:SenMWPhasNois},~\eqref{eq:NVphasSDGenJohn}, and~\eqref{eq:FiltrIntgrat}, the equivalent magnetic sensitivity due to oscillator Johnson phase noise in a multipulse NV sensor with $\tau_{\rm \pi}\,{\neq}\,0$ and $\tau_{\rm tot}\gg t_{\rm dead}$ is:
\begin{equation}
\label{eq:NoisMWLOjohsn} \eta_{\phi,J}\,{\approx}\,\frac{1}{\gamma_{\rm nv}}\,\sqrt{\frac{\pi\,f_c\,(N+1)\,{\times}\,10^{\mathcal{L}_{\rm min}/10}}{2\,\tau_{\rm tot}}}.
\end{equation}

\begin{figure}[t]
\includegraphics[width=\columnwidth]{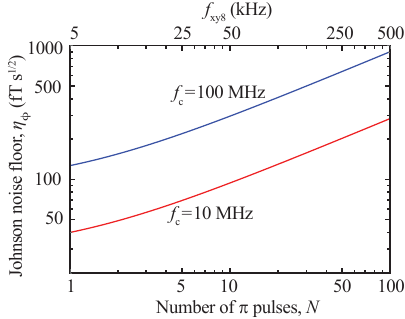}\hfill
\caption{\textbf{Oscillator Johnson phase noise limit: pulsed.} Equivalent magnetic sensitivity in the oscillator-Johnson-phase-noise limit for a CPMG or XY8 multipulse sequence, $\eta_{\phi,J}(N)$, calculated using Eq.~\eqref{eq:NoisMWLOjohsn}, with $\tau_{\rm tot}=50~{\rm \upmu s}$ and $f_c=10~{\rm MHz}$ (red) or $100~{\rm MHz}$ (blue). The first harmonic of the pulse sequence's filter function, $f_{\rm xy8}=N/(2\tau_{\rm tot})$ is shown in the top horizontal axis.}
\label{JohnsonNois}
\end{figure}

Figure~\ref{JohnsonNois} shows $\eta_{\phi,J}$ as a function of the number of $\pi$ pulses $N$, calculated using Eq.~\eqref{eq:NoisMWLOjohsn}, with $\tau_{\rm tot}=50~{\upmu s}$ and $f_c=10~{\rm MHz}$ or $100~{\rm MHz}$.

\subsection{Comment on choice of cutoff frequencies}
\label{app:cutoff}

The use of a hard cutoff frequency, $f_c$, for integrals in the filter function model, for example Eq.~\eqref{eq:FF} of the main text, is an approximation. For calculations in Fig.~\ref{fig2} in the main text, we select $f_c=0.1~{\rm GHz}$. We found that the exact choice of $f_c$ did not have much impact on our estimates of $\eta_{\phi}$ in Fig.~\ref{fig2}(a-c,f), as long as we chose $f_c\gtrsim10~{\rm MHz}$. This can be attributed to the strong roll-off of $\mathcal{L}(f)$ at offset frequencies $f\gtrsim1~{\rm MHz}$ for the generators studied here, see Fig.~\ref{fig2}(d). 

For calculations of the equivalent magnetic sensitivity in the oscillator Johnson phase noise limit in a pulsed experiment, Sec.~\ref{app:Johnsonpulse}, the results depend strongly on the choice of $f_c$, as $\eta_{\phi,J}\propto\sqrt{f_c}$, see Eq.~\eqref{eq:NoisMWLOjohsn}. In Fig.~\ref{JohnsonNois}, we plotted $\eta_{\phi,J}(N)$ for two feasible values $f_c=10~{\rm MHz}$ and $f_c=100~{\rm MHz}$. These cutoff frequencies could be due to the limited passband realized with a MW cavity. Lower values of $f_c$ may also be realized with even higher-quality-factor MW cavities, but this may come with additional challenges in pulse fidelity as the MW passband becomes comparable to the NV spin resonance spectral width.

A similar trend is found in the calculations for a continuous-wave magnetometer presented in Sec.~\ref{app:cwodmr}. The estimations of $\eta_{\rm f}$ based on $\mathcal{L}(f)$ curves of G1 and G2 depend only weakly on the choice of $f_c$ (at least for $f_c\gtrsim10~{\rm kHz}$). However, for calculations of the equivalent magnetic sensitivity in the oscillator Johnson phase noise limit the choice is important, as $\eta_{f,J}\propto\sqrt{f_c}$, see Eq.~\eqref{sigf:wh}. In Fig.~\ref{fig:Phase_noise_both}, we plot $\eta_{f,J}(\tau)$ for $f_c=1~{\rm MHz}$. This choice of $f_c$ may be justified by the assumed NV response to frequency-detuning fluctuations, but the a more accurate value of $f_c$ will depend on specific experimental conditions.

\subsection{Comment on the use of magnetic flux concentrators}
In the main text, we point out that the use of magnetic flux concentrators offers some relief from the impact of phase noise for magnetometry at appropriately low bias fields. The reason is that magnetic signals are amplified while the impact of microwave phase noise is unchanged. A magnetic flux concentrator is a ferromagnetic material that collects magnetic flux from a larger area and concentrates it into a micro-scale diamond sensor~\cite{FES2020,SHA2023,GAO2023,SIL2023}. The enhancement factor, $\epsilon$, is the ratio of the concentrated field inside the diamond to the unperturbed external field, $\epsilon=B_{\rm int}/B_{\rm ext}$. When limited by noise sources fundamental to the diamond detector (such as photon shot noise or phase noise), the sensitivity of a flux-concentrator diamond sensor to external magnetic fields is given by $\eta_{\rm ext}=\eta_{\rm int}/\epsilon$, where $\eta_{\rm int}$ is the magnetic sensitivity of the diamond without concentrators.

In Ref.~\cite{SIL2023}, a similar diamond and pulse sequence as those in this work were used, in addition to a magnetic flux concentrator that provided $\epsilon\approx300$. Also, the same generator G3 was used at carrier frequencies of $2.7{\mbox{-}}3~{\rm GHz}$. The internal magnetic sensitivity was measured to be in the range $\eta_{\rm int}=18{\mbox{-}}22~{\rm pT\,s^{1/2}}$ for detection frequencies in the range $f_{\rm xy8}=0.2{\mbox{-}}1.5~{\rm GHz}$.  The equivalent noise floor when MW were detuned was slightly lower, more consistent with the photon-shot-noise-limited prediction of ${\sim}12~{\rm pT\,s^{1/2}}$. Using the filter-function model of Eqs.\eqref{eq:FF} and \eqref{eq:etaphi} for G3 under these pulse sequences, the equivalent magnetic sensitivity due to phase noise is at the level of ${\sim}5\mbox{-}10~{\rm pT\,s^{1/2}}$. This level of noise is consistent with the quadrature difference between on-resonance and off-resonance noise floors observed in Ref.~\cite{SIL2023}. However, with flux concentrators, the overall sensor sensitivity is expected to be $\eta_{\rm ext}=\eta_{\rm int}/\epsilon\approx20~{\rm pT\,s^{1/2}}/300=70~{\rm fT\,s^{1/2}}$, which is comparable to what was observed. Thus, Ref.~\cite{SIL2023} provides evidence that, when the diamond sensor is limited by phase noise, the use of magnetic flux concentrators can still improve magnetic sensitivity down to the femtotesla level.

\section{Phase noise in a continuous wave experiment}
\label{app:cwodmr}
It is instructive to briefly look at how phase noise limits the sensitivity in a continuous-wave (cw) NV magnetometer, where both the green laser and microwave radiation are applied continuously. The microwave frequency is chosen such that a small change in spin resonance frequency corresponds to the maximum, linear change in the fluorescence rate. Assuming a Lorentzian lineshape, the fluorescence as a function of MW frequency detuning, $\Delta f(t)$, can be written as:
\begin{equation}
\label{eq:lorentzian}
    F(t)\approx F_0\left(1-C\frac{3\sqrt{3}}{4}\frac{\Delta f(t)}{\Gamma}\right),
\end{equation}
where $F_0$ is the fluorescence level, $C$ is the resonance contrast, and $\Gamma$ is the full-width-at-half-maximum resonance linewidth. Here, $\Delta f(t)$ can be caused either by a small magnetic field or by a time-dependent fluctuation in frequency due to phase noise, $\Delta f(t)= \gamma_{\rm nv} B + \delta f(t)$. 

The frequency fluctuations $\delta f(t)$ are assumed to be zero mean, with a variance that depends on the sampling interval $\tau$, given by~\cite{RUB2008}:
\begin{equation}
\label{eq:varf}
    \sigma^2_f(\tau)\approx\int_0^{f_c}f^2 S_{\phi}(f)\frac{\sin^2(\pi f\tau)}{(\pi f\tau)^2}.
\end{equation}
Here we have approximated the NV response as having a hard high-frequency cutoff at $f_c$, where the value of $f_c$ is influenced by experimental parameters such as laser intensity, microwave field strength, and the NV spin coherence times and level dynamics. The final (sinc$^2$) term in the integrand of Eq.~\eqref{eq:varf} is the filter function of the operation of averaging $\delta f(t)$ over the sampling interval $\tau$. 

Since MW frequency fluctuations affect the fluorescence signal in the same way as magnetic field noise, we can define an equivalent magnetic noise standard deviation for a single measurement over a sampling interval $\tau$ as:
\begin{equation}
\label{eq:sigmaB_cw}
\sigma_B(\tau)=\frac{\sigma_f(\tau)}{\gamma_{\rm nv}}.
\end{equation} 
Using Eq.~\eqref{eq:sigmaB_cw}, we define a sampling-interval-dependent equivalent magnetic sensitivity as:
\begin{equation}
\label{eq:etaf_cw}
\eta_f(\tau)=\sigma_B(\tau)\sqrt{\tau}.
\end{equation}

\begin{figure}[htb]
    \centering
    \includegraphics[width=\textwidth]{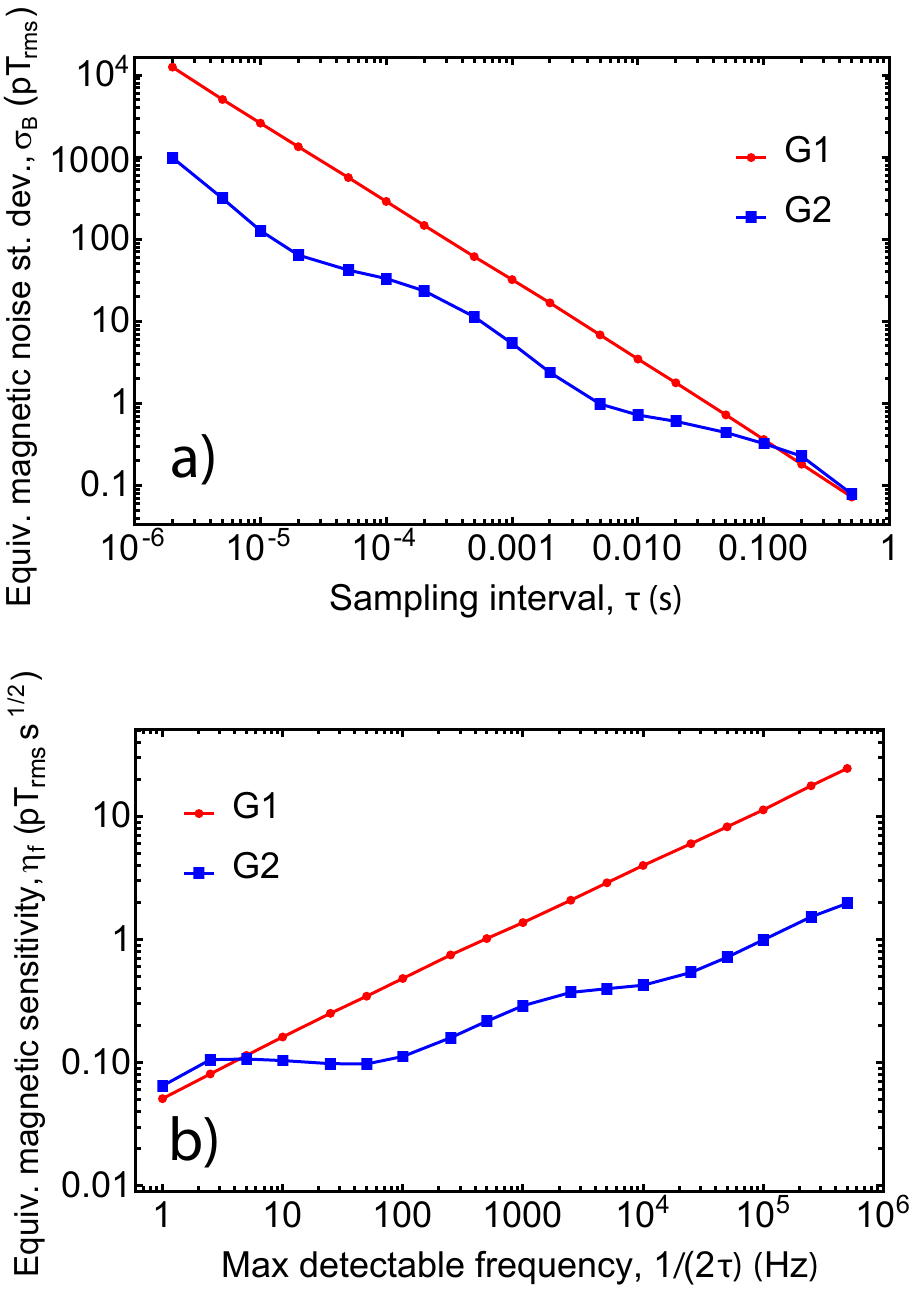}
    \caption{\textbf{Magnetic noise in a cw experiment. (a)} Estimates for equivalent magnetic noise standard deviation, $\sigma_{B}$, as a function of sampling interval $\tau$, for generators G1 and G2 at $f_{\rm G1}=2.5~{\rm GHz}$ and $f_{\rm G2}=2.1~{\rm GHz}$, respectively [see Fig.~\ref{fig2}(d) for the $\mathcal{L}(f)$ curves used]. \textbf{(b)} Estimated equivalent magnetic sensitivity $\eta_f$ as a function of magnetometer bandwidth, $1/(2\tau)$.}
    \label{fig:cw_real_gen}
\end{figure}

We can use Eqs.~\eqref{eq:varf}, \eqref{eq:sigmaB_cw}, and \eqref{eq:etaf_cw}, along with a known MW generator phase noise spectrum $S_{\phi}(f)$, to estimate $\sigma_B(\tau)$ and $\eta_f(\tau)$ in a cw experiment. Figure~\ref{fig:cw_real_gen}(a) shows the sampling rate dependence of $\sigma_B(\tau)$ for generators G1 and G2 using the phase noise spectra shown in Fig.~\ref{fig2}(d) for $f_{\rm G1}=2.5~{\rm GHz}$, $f_{\rm G2}=2.1~{\rm GHz}$, and $f_c=1~{\rm MHz}$. Figure~\ref{fig:cw_real_gen}(b) shows $\eta_f(\tau)$ as a function of the maximum detectable frequency (i.e. the magnetometer bandwidth), $1/(2\tau)$. In the cw case, the phase-noise-limited equivalent magnetic sensitivity tends to be lower than that for multipulse sequences, and it depends on the magnetometer bandwidth. However, the impact of phase noise is still important. For example, for a $1{\mbox{-}}{\rm kHz}$ bandwidth magnetometer, the equivalent magnetic sensitivity is $1.4~{\rm pT_{rms}\,s^{1/2}}$ for G1 and $0.3~{\rm pT_{rms}\,s^{1/2}}$ for G2.

\subsection{Oscillator Johnson phase noise limit: cw}
For white phase noise, such as that due to oscillator Johnson phase noise, $S_\phi(f)=S_\phi^{wh}$. 
The integral in Eq.~\eqref{eq:varf} can be done explicitly, resulting in a frequency-noise standard deviation:
\begin{equation}
\label{sigf:wh}
    \sigma_f(\tau)\approx\frac{\sqrt{2 S_\phi^{wh}f_c}}{2\pi\tau}.
\end{equation}
Inserting Eq.~\eqref{eq:varf} into Eq.~\eqref{eq:sigmaB_cw}, the equivalent magnetic-noise standard deviation for a single sampling interval $\tau$ is given by:
\begin{equation}
\label{eq:sigmaB_cwJ}
\sigma_B^{wh}(\tau)\approx\frac{\sqrt{2 S_\phi^{wh}f_c}}{2\pi\gamma_{\rm nv}\tau}.
\end{equation} 

\begin{figure}[hbt]
    \centering
    \includegraphics[width=\columnwidth]{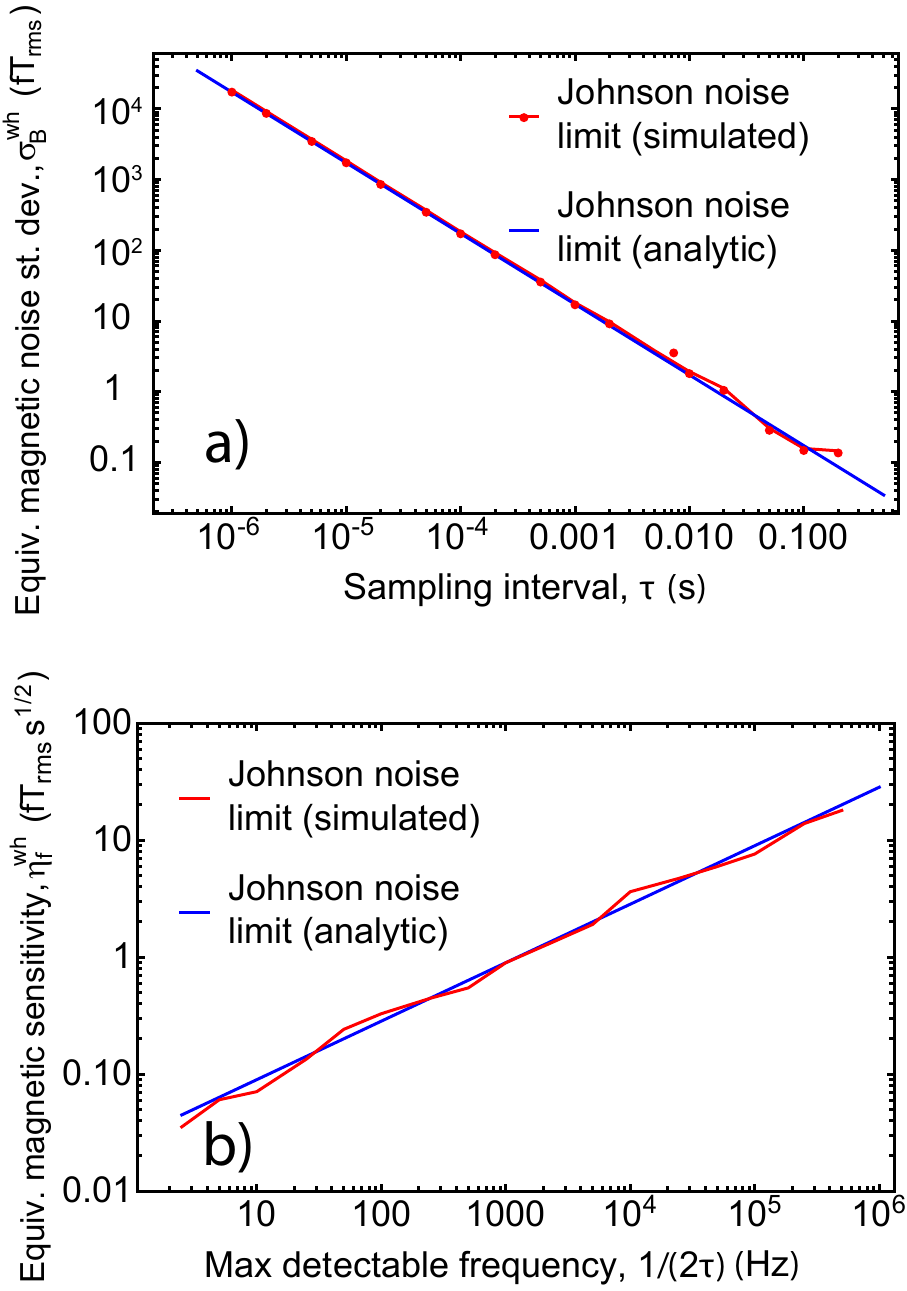}
    \caption{\textbf{Oscillator Johnson phase noise limit: cw. (a)} Equivalent magnetic noise standard deviation due to room-temperature oscillator Johnson phase noise, $\sigma_B^{wh}$, versus sampling interval $\tau$. \textbf{(b)} Equivalent magnetic sensitivity due to room-temperature oscillator Johnson phase noise, $\eta_f^{wh}$, as a function of magnetometer bandwidth, $1/(2\tau)$.}
    \label{fig:Phase_noise_both}
\end{figure}

Figure~\ref{fig:Phase_noise_both}(a) shows plots of $\sigma_B(\tau)$, using either Eq.~\eqref{eq:sigmaB_cwJ} (``analytic'') or by directly propagating numerically-simulated phase noise with Eq.~\eqref{eq:lorentzian} (``simulated''). In either case, we assume a room-temperature Johnson-noise-limited, $0~{\rm dBm}$ oscillator phase noise, $S_\phi^{wh}=2\times10^{-17.7}~{\rm rad^2/Hz}$, and $f_c=1~{\rm MHz}$. The linear dependence of $\sigma_B(\tau)$ on $\tau$ for white phase noise implies that longer sampling intervals offer superior sensitivity. However, the maximum detectable magnetic field frequency is $1/(2\tau)$, so long sampling intervals restrict the bandwidth of the magnetometer. 

Incorporating Eq.~\eqref{eq:sigmaB_cwJ} in Eq.~\eqref{eq:etaf_cw}, the equivalent magnetic sensitivity in the room-temperature, oscillator-Johnson-phase-noise limit is:
\begin{equation}
\label{eq:etaf_cwJ}
\eta_f^{wh}(\tau)=\sigma_B(\tau)\sqrt{\tau}\approx\frac{1}{\pi\gamma_{\rm nv}}\sqrt{\frac{S_\phi^{wh}f_c}{2\tau}}.
\end{equation}
Figure~\ref{fig:Phase_noise_both}(b) shows a plot of $\eta_f^{wh}(\tau)$ as a function of the maximum detectable frequency, $1/(2\tau)$. The numerically simulated values are formed by taking 1-s intervals of simulated noise, resampling by averaging over consecutive intervals of $\tau\leq0.5~{\rm s}$, and taking the maximum frequency component. The noise is then the standard deviation of 50 of these 1-s intervals. At low frequencies, the equivalent magnetic sensitivity due to Johnson phase noise is negligible, but for frequencies ${\gtrsim}100~{\rm kHz}$ the noise is ${\gtrsim}10~{\rm fT_{rms}\,s^{1/2}}$, which is above fundamental limits~\cite{TAY2008}.

\subsection{Lock-in measurements}
We anticipate that the above analysis should also hold for experiments using a lock-in amplifier. Our primary assumption is a linear relationship between the fluorescence signal and the MW carrier frequency detuning, Eq.~\eqref{eq:lorentzian}, which should also hold true for a lock-in measurement. One caveat is that, when filtering is done in a lock-in experiment, there is a minimum effective sampling interval, $\tau_{\rm min}$, that is given by $1/(2\tau_{\rm min})\approx f_{\rm enbw}\lesssim f_{\rm mod}$, where $f_{\rm enbw}$ is the effective noise bandwidth and $f_{\rm mod}$ is the modulation frequency of the lock-in experiment.

\subsection{Suppressing the impact of MW phase noise in a cw experiment}
\label{app:cwdual}
As mentioned in Sec.~\ref{sec:discuss} of the main text, we expect that the dual-resonance scheme, first described in Ref.~\cite{WOJ2018} (see also Ref.~\cite{FES2020}), should be effective at suppressing frequency noise in cw optically-detected magnetic resonance measurements. Consider the MW mixer schematic presented in Fig.~\ref{DQSchem}(a) with the ``V'' level diagram on the left side of Fig.~\ref{DQSchem}(b). The MW Rabi frequency and optical pump rate are assumed to be weak enough, relative to the spin resonance linewidth, to avoid populating coherent dark states~\cite{SHI2022}. If the output frequencies of the mixer are tuned to drive opposing sides of the $f_{\pm}$ resonances, the impact of the frequency noise of the main oscillator, with carrier frequency $D=2.87~{\rm GHz}$, on the NV signal should cancel. The remaining frequency noise from the local oscillator does not cancel, but its carrier frequency is much smaller, ${\sim}\gamma_{\rm nv}B_0\ll D$. If the frequency noise scales linearly with carrier frequency, the result should be an improvement in the frequency-noise-limited sensitivity by a factor ${\sim}D/(\gamma_{\rm nv}B_0)$.

\section{Time-domain model of phase-noise-limited sensitivity}
\label{app:tdomainmodel}

In this section, formulas are derived for the equivalent magnetic field noise resulting from MW pulse sequences with white and random-walk phase noise. We focus on pulse sequences using a series of $N$ resonant MW $\pi$ pulses with uniform spacing, such as CPMG and XY pulse sequences. However, a similar analysis can be applied to a wide variety of pulse sequences. The MW pulses are in resonance with only one of the NV $f_{\pm}$ spin transitions such that the spin dynamics can be described by an effective spin-1/2 system. We assume all pulses are instantaneous, resonant pulses and there are no additional time-varying fields.

\begin{figure}[htb]
\includegraphics[width=\columnwidth]{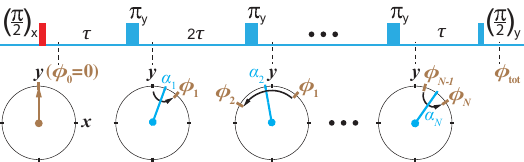}\hfill
\caption{\textbf{Propagation of spin state rotations due to MW phase errors.} A graphical representation of the model for a spin-$1/2$ qubit under a multipulse ($N~\pi$-pulses) sequence in the absence of time-varying magnetic fields. All pulses are assumed to be ideal, resonant pulses of negligible length. The first MW $\pi/2$ pulse rotates the NV spin to lie along the $y$-axis (Bloch vector angle: $\phi_0=0$), where the $x$ and $y$ axes are defined by the MW carrier's phase during this pulse. For subsequent MW $\pi$ pulses, the spin state is reflected about a rotated axis of angle $\alpha_i$ with respect to the $y$-axis, where $\alpha_i$ is the MW phase error. The spin state's phase after the $i$'th pulse is $\phi_i=2\alpha_i-\phi_{i-1}$, see Eq.~\eqref{eq:PhasEachPi}.
}
\label{BlockSphNois}
\end{figure}

Figure~\ref{BlockSphNois} shows an example CPMG pulse sequence along with the associated phase of the NV spin state on the equator of the Bloch sphere. NV centers are initially polarized along the $z$-axis of the Bloch sphere. A MW $\pi$/2 pulse along the $x$-axis rotates the NV electron spin to lie along the $y$-axis. The phase of this first $\pi/2$ pulse defines the coordinates of the Bloch sphere for all subsequent pulses. At this point, the NV phase is defined to be $\phi_0=0$ with respect to the $y$-axis, and subsequent errors in the microwave phase are defined relative to this axis. Next, $N$ resonant $\pi$-pulses, spaced a time $2\tau$ apart, are applied. While the $\pi$ pulses are applied approximately along the $y$-axis, each pulse has a small phase error, $\alpha_i$, with respect to the $y$-axis. The effect of each $\pi$ pulse is to apply a $\pi$ rotation of the NV spin state about the axis defined by the MW phase error, $\alpha_i$. In the absence of external magnetic signals, the phase of the NV spin state $\phi_i$ after the $i$'th $\pi$-pulse can be written as: 
\begin{equation}
\label{eq:PhasEachPi}
\phi_i~{=}~2\,(\alpha_i\,{-}\,\phi_{i-1})+\phi_{i-1}=2\,\alpha_i-\phi_{i-1},
\end{equation}
where $\phi_{i-1}$ is the phase of the spin state prior to the pulse. The total phase accumulated by the NV spin state, $\phi_{\rm tot}$, at the end of the pulse sequence (Fig.~\ref{BlockSphNois}) is given by:
\begin{equation}
\label{eq:PhasTotCPMG}\phi_{\rm tot}~{=}~-\alpha_f+\phi_N\,~{=}~\,-\alpha_f+\sum_{i=1}^{N}(-1)^{N-i}\,2\,\alpha_{i}\,,
\end{equation}
where $\alpha_f$ is the MW phase error of the final $\pi/2$ pulse.

\subsection{White microwave phase noise}
Equation~\eqref{eq:PhasTotCPMG} shows how MW phase errors lead to an error in the NV spin state's phase. We now consider the case when the MW phase errors, $\alpha_i$, are normally distributed, as is the case for white noise. White phase noise is especially prominent at higher MW carrier offset, as seen in Fig.~\ref{fig2}(d). A fundamental constraint on the phase noise performance of MW generators is due to Johnson noise in the MW oscillator which has a white noise character.

The variance of the NV total phase error, ${\rm Var}(\phi_{\rm tot}$), is the sum of the variances of each term on the right side of Eq.~\eqref{eq:PhasTotCPMG}:
\begin{equation}
\label{eq:Var_PhasTotCPMG} {\rm Var}({\phi_{\rm tot}})~{=}~{\rm Var}(\alpha_f)+\sum_{i=1}^{N} {\rm Var}(2\,\alpha_i).
\end{equation}
In the case of white noise, the variance of the MW phase errors, ${\rm Var}(\alpha_{f,i})$, is independent of the temporal spacing between the pulses and only depends on the standard deviation $\sigma_{\rm wh}$ of the MW phase error in each pulse. In other words:
\begin{equation}
\label{eq:Var_EachPuls} {\rm Var}(\alpha_i)~{=}~{\rm Var}(\alpha_f)~{=}~\sigma_{\rm wh}^{2}.
\end{equation}
Inserting Eq.~\eqref{eq:Var_EachPuls} in Eq.~\eqref{eq:Var_PhasTotCPMG}, the total variance becomes:
\begin{equation}
\label{eq:TotVarWhit} {\rm Var}({\phi_{\rm tot}})~{=}~4\,\sigma_{\rm wh}^2\,(N+1/4),
\end{equation}
and the standard deviation of the total phase error of the NV spin state, $\sigma_{\rm \phi}$, is:
\begin{equation}
\label{eq:SDTotWhit} \sigma_{\rm \phi}~{=}~\sqrt{{\rm Var}({\phi_{\rm tot}})}~{=}~2\,\sigma_{\rm wh}\,\sqrt{N+1/4}.
\end{equation}
Inserting Eq.~\eqref{eq:SDTotWhit} in Eq.~\eqref{eq:etaphi}, using $\tau_{\rm tot}=2N\tau$ and assuming $N\gg1$, the white-phase-noise-limited equivalent magnetic sensitivity can be written as:
\begin{equation}
\label{eq:SenMWPhasWhit}
\eta_{\rm wh}\,{\approx}\,\frac{\sigma_{\rm wh}\,\sqrt{N(1+t_{\rm dead}/\tau_{\rm tot})}}{2\,\gamma_{\rm nv}\sqrt{\tau_{\rm tot}}}\,{=}\,\frac{\sigma_{\rm wh}}{\gamma_{\rm nv}}\,\sqrt{\frac{f_{\rm xy8}}{2\,\delta}},
\end{equation}
where $f_{\rm xy8}=1/(4\,\tau)$ is the central frequency of the pulse sequence's filter function (we use the subscript $\rm xy8$ here, but the same definition applies for CPMG sequences) and $\delta=\tau_{\rm tot}/(\tau_{\rm tot}+t_{\rm dead})$ is the duty cycle of the NV readouts.

\subsection{Random-walk microwave phase noise}
If the MW phase errors ($\alpha_i$) are dominated by random-walk noise, the relative MW phase error accumulates. The phase error of the $i$'th pulse is given by: 
\begin{equation}
\label{eq:FramRotRW} \alpha_i\,=\,\alpha_{i-1}+\Delta_i,
\end{equation}
where $\Delta_i$ is the MW phase displacement accrued during the time interval between pulses $i-1$ and $i$. Similar to the white noise case, we suppose that the first $\pi$/2 pulse in Fig.~\ref{BlockSphNois} is perfect, $\alpha_0\,{=}\,0$. Inserting Eq.~\eqref{eq:FramRotRW} into Eq.~\eqref{eq:PhasTotCPMG}, the total phase error of the NV spin state for even and odd numbers of the applied $\pi$-pulses can be written as: 
\begin{equation}
\label{eq:PhasTotRWEvntemp}
\phi_{\rm tot}=
\begin{cases}
\,\,-\alpha_f+\displaystyle\sum_{i=1}^{N/2}\,
  2\,\Delta_{2i} &  
  \text{$N$ even} \\
  \\
  \,\,-\alpha_f+\displaystyle
  \sum_{i=1}^{(N+1)/2}
  \,2\,\Delta_{2i-1}
   & \text{$N$ odd.}
\end{cases}
\end{equation}
\\
This can be further reduced to:
\begin{equation}
\label{eq:PhasTotRWEvn}
\phi_{\rm tot}=
-\Delta_f+\displaystyle\sum_{i=1}^{N}(-1)^{N-i}\,\Delta_{i}.
\end{equation}

In Eq.~\eqref{eq:PhasTotRWEvn}, $\Delta_1$ and $\Delta_{f}$ are special cases where the phase displacement is accrued over an interval $\tau$ (all other $\Delta_i$ accrue over an interval $2\tau$), and their variances can be assumed to be the same. Specifically:
\begin{equation}
\label{eq:DeltaPiHafRW}{\rm Var}(\Delta_1)~{=}~{\rm Var}(\Delta_{f})~{=}~\sigma_{\rm rw}^{2}\,R_{\rm samp}\,\tau.
\end{equation}
Here we have modeled the random-walk MW phase evolution as a set of discrete jumps in MW phase, where $\sigma_{\rm rw}$ is the standard deviation of the phase jumps and $R_{\rm samp}\gtrsim1/(2\tau)$ is the jump sampling rate. 

Because the $\pi$ pulses are evenly spaced in time by $2\tau$ (see Fig.~\ref{BlockSphNois}), the variance of the random-walk MW phase displacement of all other $\Delta_i$ ($i=2,3,...N$) is two times larger than those in Eq.~\eqref{eq:DeltaPiHafRW}:
\begin{equation}
\label{eq:DeltaPiRW} {\rm Var}(\Delta_2)\,{=}\,{\rm Var}(\Delta_3)\,{=}\,...\,{=}\,{\rm Var}(\Delta_N)\,{=}\,\sigma_{\rm rw}^{2}\,R_{\rm samp}\,2\,\tau.
\vspace{3 mm}
\end{equation}

Using Eqs.~\eqref{eq:DeltaPiHafRW} and Eq.~\eqref{eq:DeltaPiRW} with Eq.~\eqref{eq:PhasTotRWEvn}, the variance of the phase of the NV spin state due to MW random-walk phase noise becomes:
\begin{equation}
\label{eq:VarPhasTotRW}{\rm Var}(\phi_{\rm tot})\,{=}\,\sigma_{\rm rw}^{2}\,2\,N\,\tau\,R_{\rm samp},
\end{equation}
and its standard deviation is:
\begin{equation}
\label{eq:SDPhasTotRW}\sigma_{\phi}\,{=}\,\sigma_{\rm rw}\,\sqrt{2\,N\,\tau\,R_{\rm samp}}\,{=}\,\sigma_{\rm rw}\,\sqrt{\tau_{\rm tot}\,R_{\rm samp}}.
\end{equation}
Inserting Eq.~\eqref{eq:SDPhasTotRW} into Eq.~\eqref{eq:etaphi} of the main text, the random-walk-phase-noise-limited equivalent magnetic sensitivity is given by:
\begin{equation}
\label{eq:SenNoisRW} \eta_{\rm rw}\,\approx\,\frac{\sigma_{\rm rw}}{4\,\gamma_{\rm nv}}\sqrt{\frac{R_{\rm samp}}{\delta}}.
\end{equation}

\subsection{Extension to XY8-$N_{\rm r}$ pulse sequences}
In the case of the XY8-$N_r$ pulse sequence used in this study, the total number of the applied $\pi$-pulses is $N=8N_{\rm r}$. The total NV spin-state phase accumulation due to MW phase errors is given by:
\begin{equation}
\label{eq:PhasTotXY8}\phi_{\rm tot}~{=}~-\alpha_f+\sum_{i=1}^{N}(-1)^{i}\,2\alpha_{i},
\end{equation}
which is identical to the expression for $N$ even derived for the CPMG sequence, Eq.~\eqref{eq:PhasEachPi}. Here the MW phase errors $\alpha_i$ are with respect to the desired phase. For XY8 sequences, the desired phase alternates between $0\degree$ and $90\degree$, but it does so in such a way that there is no final phase accumulation in the absence of spin precession or phase errors. Thus the expressions derived above for $\sigma_{\phi}$ [Eqs.~\eqref{eq:SDTotWhit}, \eqref{eq:SDPhasTotRW}], and $\eta_{\phi}$ [Eqs.~\eqref{eq:SenMWPhasWhit}, \eqref{eq:SenNoisRW}] are also valid for the XY8-$N_{\rm r}$ sequence.

\section{Phase noise in double-quantum measurement}
\label{app:dq}

In the case of double-quantum measurements, two microwave fields with the same amplitude are simultaneously applied at both $f_{\pm}$ transition frequencies. Here we assume the bias magnetic field is small, $B_0\ll D/\gamma_{\rm nv}$.

\begin{figure}[htb]
\includegraphics[width=0.95\columnwidth]{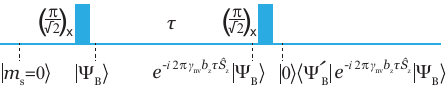}\hfill
\caption{\textbf{Double-quantum Ramsey sequence.} Microwave pulses are formed from two tones resonant with both $f_{\pm}$ transitions. The first MW pulse transfers the NV population from $\ket{m_{\rm s}\,{=}\,0}$ state to the "bright" coherent superposition state $\ket{\Psi_{\rm B}}$, Eq.~\eqref{eq:BrightCPT}. The state then freely evolves under a small magnetic field $b_{\rm z}$ for a time $\tau$. A second MW pulse transfers the bright state $\ket{B'}$ back to $\ket{m_{\rm s}\,{=}\,0}$ state. Note that $\ket{B}\neq\ket{B'}$ if the relative phase of the MW tones changes, see Eqs.~\eqref{eq:SzRamsy}, \eqref{eq:BrightCPT}.}
\label{Ramsey2ToneSeq}
\end{figure}

Figure~\ref{Ramsey2ToneSeq} depicts a double-quantum Ramsey pulse sequence. The first two-tone resonant MW pulse transfers the NV population from $\ket{m_{\rm s}\,{=}\,0}$ state to the "bright" coherent superposition state $\ket{\Psi_{\rm B}}$ given by~\cite{SCU1997,SHA2010}:
\begin{equation}
\label{eq:BrightCPT}
\ket{\Psi_{\rm B}}\,{=}\,\frac{1}{\sqrt{2}}\,e^{-i\frac{\alpha_1+\alpha_2}{2}}\bigl[e^{i\,\frac{\Delta\alpha}{2}}\ket{-1}\,{+}\,e^{-i\,\frac{\Delta\alpha}{2}}\,\ket{+1}].
\vspace{3 mm}
\end{equation}
In Eq.~\eqref{eq:BrightCPT}, $\alpha_1$ and $\alpha_2$ are the phases of the two resonant MW tones, and $\Delta\alpha=\alpha_2-\alpha_1$. 

If the second two-tone pulse, with MW phases $\alpha_1'$ and $\alpha_2'$, is applied in a time interval $\tau$ (Fig.~\ref{Ramsey2ToneSeq}), it will project the NV population back into the $\ket{0}$ state with a probability given by:
\begin{equation}
\label{eq:SzRamsy}
|\bra{\Psi_{\rm B}'}^{-i2\pi\gamma_{\rm nv}b_z\tau\hat{S_{\rm z}}}\ket{\Psi_{\rm B}}|^{2}{=}\cos^{2}{(\frac{\Delta\alpha'{-}\Delta\alpha}{2}{-}2\pi\gamma_{\rm nv}b_z\tau)},
\vspace{3 mm}
\end{equation}
where $\Delta\alpha'\,{=}\,\alpha_2'\,{-}\,\alpha_1'$. In Eq.~\eqref{eq:SzRamsy}, $b_z\ll \Omega_{\rm R}/\gamma_{\rm nv}$ is a small magnetic field to be detected that has a negligible effect on the NV coherent bright states, Eq.~\eqref{eq:BrightCPT}. If this magnetic field is very small, $2\pi \gamma_{\rm nv} b_z\tau\ll1$, then $\Delta \alpha'$ can be offset by $90\degree$ to ensure the probability in Eq.~\eqref{eq:SzRamsy} depends linearly on $b_z$.

\begin{figure}[htb]
\includegraphics[width=\columnwidth]{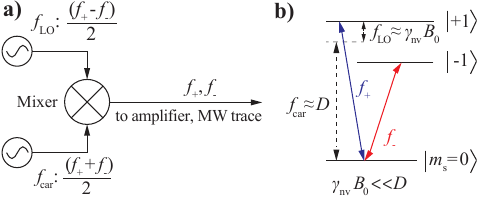}\hfill
\caption{\textbf{Double-quantum and dual-resonance schemes.} \textbf{(a)} Electronic schematic for generating dual-tone MW frequencies in resonant with NV spin transitions $f_{\pm}$~\cite{MAM2014,FES2020}. $f_{\rm LO}$: Frequency of local oscillator, $f_{\rm car}$: Frequency of carrier generator. \textbf{(b)} Ground-state NV spin level diagram for the ``V'' scheme ($B_{0}\ll D/{\rm \gamma_{\rm nv}}$).}
\label{DQSchem}
\end{figure}

Figure~\ref{DQSchem}(a) shows a method to generate the two MW tones that minimizes the impact of phase noise~\cite{MAM2014,FES2020}. Figure~\ref{DQSchem}(b) shows the associated NV level diagrams in the low-field regime. By mixing a lower-frequency signal generator as the local oscillator [$f_{\rm LO}=(f_+-f_-)/2\approx\gamma_{\rm nv} B_0$] and a higher-frequency signal generator as the carrier [$f_{\rm car}=(f_++f_-)/2\approx D$], both the NV spin transitions with frequencies $f_{\rm \pm}$ can be driven simultaneously. Moreover, the phase noise of the higher-frequency $f_{\rm car}$ tone has no impact on the magnetometer signal [see Eq.~\eqref{eq:SzRamsy} and Sec.~\ref{app:mixer}], leaving only the phase noise due to $f_{\rm LO}$. If the phase noise scales linearly with MW carrier frequency, this scheme can suppress the impact of MW phase noise by a factor ${\sim}D/\gamma_{\rm nv}B_0$ in low fields ($\gamma_{\rm nv}B_0\ll D$).

This measurement protocol is analogous to how high-precision atomic spectroscopy of 3-level atoms is conducted in the optical regime with noisy lasers~\cite{THO1983}. At very low applied field $B_0\lesssim\Omega_R/\gamma_{\rm nv}$, where $\Omega_R$ is the MW Rabi frequency, a single MW tone can be used with a double-quantum pulse sequence~\cite{FAN2013}, potentially eliminating the first-order impact of MW phase noise entirely. At high magnetic field, $B_0>D/\gamma_{\rm nv}$, this method does not appear to lead to a phase-noise cancellation, and alternative methods should be pursued.

%

\bibliographystyle{apsrev4-2}

\end{document}